\documentclass[twocolumn,aps,pra,superscriptaddress, 10pt]{revtex4-2} 
\usepackage{graphicx}  
\usepackage{dcolumn}   
\usepackage{bm}        
\usepackage{amssymb}   
\usepackage{amsmath}   
\usepackage[version=4]{mhchem}
\usepackage{siunitx}
\usepackage{natbib}
\usepackage{braket}
\usepackage[colorlinks=true, linkcolor=blue, urlcolor=blue, citecolor=blue]{hyperref} 
\usepackage{float}
\usepackage{xcolor}
\usepackage[T1]{fontenc}

\usepackage{booktabs}

\begin{document}

\title{Rapid Charge Stability Diagram Generation from Device-level Modeling of Semiconductor Quantum Dots}

\author{Ron Nodel}
\email{rnodel@g.ucla.edu}
\affiliation{Center for Quantum Science and Engineering, University of California, Los Angeles, 90095, California, USA}
\affiliation{Department of Electrical and Computer Engineering, University of California, Los Angeles, 90095, California, USA}

\author{David W. Kanaar} 
\affiliation{Center for Quantum Science and Engineering, University of California, Los Angeles, 90095, California, USA}
\affiliation{Department of Electrical and Computer Engineering, University of California, Los Angeles, 90095, California, USA}

\author{Connor Nasseraddin} 
\affiliation{Center for Quantum Science and Engineering, University of California, Los Angeles, 90095, California, USA}
\affiliation{Department of Electrical and Computer Engineering, University of California, Los Angeles, 90095, California, USA}
\affiliation{Physics and Astronomy Department, University of California, Los Angeles, 90095, California, USA}

\author{Tim J. Wilson}
\affiliation{Physics and Astronomy Department, University of California, Los Angeles, 90095, California, USA}
\affiliation{National Institute of Standards and Technology, Gaithersburg, MD 20899, USA}
\affiliation{Joint Quantum Institute, University of Maryland, College Park, MD 20742, USA.}

\author{Hong-Wen Jiang}
\affiliation{Center for Quantum Science and Engineering, University of California, Los Angeles, 90095, California, USA}
\affiliation{Physics and Astronomy Department, University of California, Los Angeles, 90095, California, USA}

\author{Jason R. Petta}
\affiliation{Center for Quantum Science and Engineering, University of California, Los Angeles, 90095, California, USA}
\affiliation{Department of Electrical and Computer Engineering, University of California, Los Angeles, 90095, California, USA}
\affiliation{Physics and Astronomy Department, University of California, Los Angeles, 90095, California, USA}

\author{Chris Anderson} 
\affiliation{Center for Quantum Science and Engineering, University of California, Los Angeles, 90095, California, USA}
\affiliation{Department of Mathematics, University of California, Los Angeles, 90095, California, USA}

\author{Mark F. Gyure} 
\affiliation{Center for Quantum Science and Engineering, University of California, Los Angeles, 90095, California, USA}
\affiliation{Department of Electrical and Computer Engineering, University of California, Los Angeles, 90095, California, USA}

\date{\today}

\begin{abstract}
Self-consistent Schrödinger-Poisson calculations are a powerful tool for predicting the behavior of layered semiconductor quantum dot devices. However, characterization of charge stability diagrams through fully simulated gate-voltage sweeps is computationally expensive.
Combining a Multi-Domain Multi-Model (MDMM) approach with an automated tuning routine, we identify gate voltages associated with selected charge configurations. This small set of self-consistent simulations can be augmented with Full Configuration Interaction (FCI) energy calculations to extract charging energies, lever arms, and interdot Coulomb interactions to directly parameterize a Hubbard model for rapid charge stability diagram generation. For an Intel Tunnel Falls Si/SiGe device, we demonstrate the Hubbard model's ability to reproduce charge stability diagrams at a fraction of the computational cost in comparison to voltage bias sweeps. We further compare the simulated diagrams to experimental data and demonstrate qualitative agreement. Our result represents a step towards predictive digital twin models for semiconductor quantum dot devices. Finally, we apply this workflow towards lever arm engineering in a second device, demonstrating that the method extends to multiple architectures.
\end{abstract}

\maketitle

\section{Introduction}
Semiconductor quantum dots have emerged as a promising platform for quantum computing due to their scalability and compatibility with the existing semiconductor industry \cite{zwerverQubitsMadeAdvanced2022,zwanenburgSiliconQuantumElectronics2013, vandersypenQuantumComputingSemiconductor2019, burkardSemiconductorSpinQubits2023}. In these systems, the spin degree of freedom of an electron in a quantum dot provides a natural two-level system that can serve as the qubit, while maintaining long coherence times due to weak coupling of the spin to the environment. Gate-defined quantum dots allow these spins to be controlled using electrostatic gate electrodes, enabling natural integration with field-effect transistors for hybrid quantum-classical operation \cite{lossQuantumComputationQuantum1998,zwanenburgSiliconQuantumElectronics2013,vandersypenQuantumComputingSemiconductor2019}. However, the design and operation of such devices remains challenging. Device performance is sensitive to the electrostatic potential and heterostructure parameters, making it difficult to optimize device designs within fabrication constraints and experimental tuning procedures \cite{zwerverQubitsMadeAdvanced2022}.

A key step in understanding and operating gate-defined quantum dot devices is analyzing charge stability diagrams, which map applied gate voltages to stable charge configurations of a device \cite{vanderwielElectronTransportDouble2002}. Experimentally, these diagrams provide insight into the charging energies, lever arms, and interdot interactions that govern device behavior. Predicting these diagrams from device design is a challenging simulation problem due to the need to accurately capture the effects of multiple gates, charge reservoirs, and heterostructure layers.

Predictive modeling of quantum dot devices typically requires solving the Schrödinger-Poisson (SP) equations self consistently in order to capture both the electrostatic environment and the quantum-mechanical influence of the electrons. Although high accuracy single-particle simulations can describe confinement of electrons trapped in a realistic gate potential, they often fail to capture the influence of charge reservoirs on device behavior. Consequently, fully self-consistent simulations are often necessary to capture relevant physics, but performing such simulations across the full gate-voltage space of a charge stability diagram is computationally expensive. 

Several previous works have employed simplified models, such as capacitance models \cite{vanderwielElectronTransportDouble2002,zwanenburgSiliconQuantumElectronics2013,cheNeuroQDLearningBasedSimulation2025, hansonSpinsFewelectronQuantum2007}
or effective Hubbard models \cite{burkardCoupledQuantumDots1999, yangGenericHubbardModel2011, wangQuantumTheoryChargestability2011,merinoSimulatedChargeStability2025,foulkTheoryChargeStability2024,wangEfficientCharacterizationDouble2024}, to generate charge stability diagrams more efficiently. While these approaches capture the key physics underlying the charge stability diagrams, they do not fully incorporate the electrostatics of a realistic device, including asymmetric gate geometries and nearby charge reservoirs. Conversely, other studies have parameterized Hubbard models directly from experimental data, allowing device behavior to be reproduced phenomenologically but without a direct connection to underlying device physics \cite{yangGenericHubbardModel2011}. 


Such a model is valuable beyond predicting charge stability diagrams. Gate-defined semiconductor spin qubits provide a natural platform for simulating Fermi-Hubbard Hamiltonians \cite{staffordCollectiveCoulombBlockade1994,byrnesQuantumSimulationFermiHubbard2008,hensgensQuantumSimulationFermi2017,salfiQuantumSimulationHubbard2016,wangExperimentalRealizationExtended2022}. Realizing a target Hubbard model experimentally therefore requires a clear correspondence between the gate architecture and the resulting Fermi-Hubbard parameters. Extracting these parameters from a realistic electrostatic model would allow gate designs to be explored efficiently and steered toward specific quantum simulation targets.

In this work, we target these goals by combining self-consistent SP simulations with the rapid parametrization of an effective Hubbard model in order to generate charge stability diagrams. 
Using the Multi-Domain Multi-Model (MDMM) framework within the Modeling and Simulation for Quantum Exploration (MaSQE) simulation suite, we employ an automated tuning procedure to identify a small set of gate voltage configurations corresponding to specific charge states of the device. 
From these simulations and their corresponding FCI calculations, we extract the effective parameters of a Hubbard model, which can then be used to rapidly generate charge stability diagrams across voltage space. Producing charge stability diagrams directly from the gate layout and heterostructure is a step toward predictive digital twins for semiconductor quantum devices. 

We validate this method by comparing the Hubbard model charge stability diagrams with self-consistent gate voltage bias-sweep simulations and experimental measurements. The FCI-Hubbard model and SP approach both show qualitative agreement with experimental data from an Intel Tunnel Falls device. Quantitative agreement is observed with the FCI-Hubbard model following a modest correction to the assumed dielectric stack. In addition, we demonstrate how these codes can be used to optimize gate designs prior to fabrication. Together, these results indicate that essential electrostatic physics can be captured using a small number of self-consistent simulations. 

We demonstrate our approach for double quantum dot systems as charge stability diagrams are commonly represented in this manner, though the method can be directly extended to larger architectures. In Section \ref{sec:Methods} we describe the MaSQE simulation framework and the MDMM SP solver used to simulate the device electrostatics. We then outline the procedure used to extract the parameters of the effective Hubbard model. In Section \ref{sec:results} we present results for both an Intel Tunnel Falls Si/SiGe device and an overlapping gate device \cite{wilsonFastSensitiveReadout2026}, comparing the Hubbard model predictions to bias-sweep simulations and experimental data. Finally, Section \ref{sec:conclusion} summarizes the results and discusses possible future directions for extending this framework.

\section{Methods}\label{sec:Methods}
In this section, we first summarize the MDMM approach to finding the solution of the Schrödinger-Poisson equations using MaSQE, and then detail how we use it to parameterize an asymmetric Hubbard model for rapid charge stability diagram generation. 
\subsection{Multi-Domain Multi-Model Schrödinger-Poisson} \label{MDMM}
The self-consistent Schrödinger-Poisson (SP) solver enhanced with a Multi-Domain Multi-Model (MDMM) approach \cite{andersonHighOrderAccurate2021, andersonEfficientSolutionSchroedinger2009, andersonFourierWachspressMethod2005} functions by finding a self-consistent solution to both Poisson's equation
\begin{equation} \label{poisson}
\Delta_{0} \Phi_{\text{var}}  = \sum_i-\rho_i
\end{equation}
and an effective Schrödinger density operator
\begin{equation}\label{schro}
\rho_i = {S_i}(\Phi_{\text{fixed}}+\Phi_{\text{var}} )
\end{equation}
across $i$ spatial subdomains. $\rho_i$ denotes the charge density in subdomain $i$, with the total device charge density given by the sum over all subdomains.
A subdomain-specific Schrödinger density operator, or ``charge filling model", ${S_i}$, can be chosen to approximate the charge density with varying accuracy across different subdomains, as discussed later in this section. $\Delta_{0}$ is the discrete Poisson operator with homogeneous boundary conditions and $\Phi_{\text{fixed}}$ is the potential due to the gate voltages applied to the metallic gates on top of the device. $\Phi_{\text{var}}$ is the variable potential induced by the sum over the charge in each subdomain, determined from the Schrödinger operator in each step. As described in Ref. \cite{andersonEfficientSolutionSchroedinger2009}, the self-consistent solution of equations \hyperref[poisson]{(1)} and \hyperref[schro]{(2)} is obtained by characterizing $\Phi_{\text{var}}$ as the solution of a time-dependent ordinary differential equation and evolving that equation to steady state. Once $\Phi_{\text{var}}$ is determined, evaluating \hyperref[schro]{(2)} yields the corresponding self-consistently determined charge densities. 



\begin{figure}
        \centering
            \includegraphics[width=\linewidth,trim={14bp 2bp 4bp 2bp},clip]{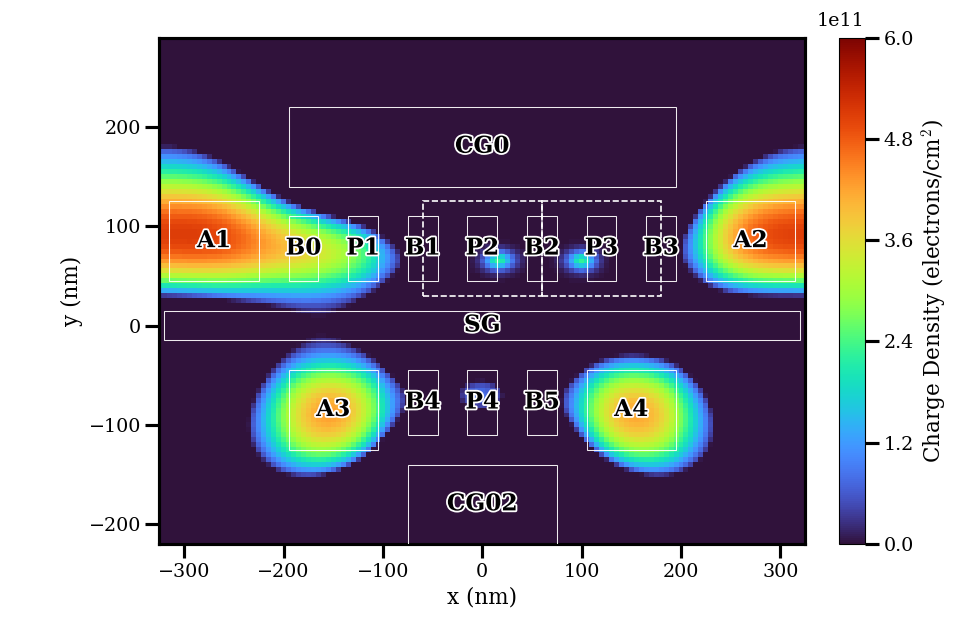}
        \caption{The approximate Tunnel Falls gate geometry and calculated charge density. The gate names are overlayed, with the MDMM subdomains shown by dashed lines around P2 and P3. The two-dimensional charge density associated with one electron in both P2 and P3 (1,1) is shown in the color map and is found using the autotuning procedure described in Section \ref{MDMM}, integrated over the quantum well thickness. This is a triple quantum dot tuned up as a double quantum dot, with P1 and B1 acting as accumulation gates, following standard experimental tune-up. Sensor accumulation gates are modeled as smaller for computational efficiency and have minimal effect on the P2-P3 charge stability diagram.}
        \label{fig:intel_1-1_geometry}
\end{figure}

We first examine the Intel triple quantum dot Tunnel Falls device \cite{marcksValleySplittingCorrelations2025,yooDirectlyVisualizingEnergy2026, george12SpinQubitArraysFabricated2025, neyensProbingSingleElectrons2024}, with approximate gate geometry shown in Fig.~\ref{fig:intel_1-1_geometry}. Since the actual design parameters are proprietary, we approximate the Tunnel Falls gate geometry and assume heterostructure parameters with a 60 nm gate pitch \cite{neyensProbingSingleElectrons2024}. These metal gates define part of the electrostatic boundary conditions at the top of the heterostructure. In the MDMM approach, we partition the device into user-defined subdomains in which a refined calculation can be applied. Within these subdomains, a higher spatial resolution and a full three-dimensional quantum-mechanical model can be employed. In this work, we specify two such subdomains corresponding to the two gate-defined quantum dots beneath P2 and P3. The subdomains are indicated by the dashed outlines in Fig.~\ref{fig:intel_1-1_geometry} and we focus on computing their charge occupancies, $(n_1,n_2)$, where $n_i$ is the number of free electrons in dot i.

Standard SP calculations map gate voltages to charge configurations. 
In contrast, we build on the MDMM approach with an automated tuning (autotuning) routine that solves the inverse problem. This approach allows for specified charge configurations to be set in each subdomain, with gate voltages adjusted iteratively until a self-consistent solution is determined. This solution corresponds to when a specified number of single particle states in an occupied dot are slightly below the Fermi level ($-1\pm0.001$ meV) and the remaining single particle states have an energy above the Fermi level. However, charge states with multiple electrons require a Full Configuration Interaction (FCI) calculation in order to properly capture the energy associated with the Coulomb interaction and we perform these calculations following the method in Ref. \cite{andersonHighOrderAccurate2021}. Solving this inverse problem is especially important in parameterizing a Hubbard model, which requires the voltages corresponding to a charge configuration. In subdomains where the autotuning procedure is used to determine gate biases for a specified charge occupation, a fully quantum single particle Schrödinger operator is used for the associated $S_i$. 

For the bias-sweep simulations, we utilize a more approximate continuous Thomas--Fermi charge filling model for $S_i$ \cite{andersonEfficientSolutionSchroedinger2009}. In this model, the growth direction uses a one-dimensional effective-mass quantum calculation for the electrons with a lateral semi-classical density of states model. Details on the variation between results using the fully quantum three-dimensional model and the charge filling model can be found in appendix \ref{appendix_full_3qd}. To convert the semi-classical continuous model to a discrete one, we threshold the filling by rounding down the charge obtained to a whole number. 

As an example of an autotuned self-consistent SP solution, Fig.~\ref{fig:intel_1-1_geometry} shows the charge density for the case where the voltages are tuned so that a single electron is induced under both P2 and P3 (1,1). 
Since the Hubbard model is parameterized as a function of energy, an FCI calculation is the most accurate way to calculate total energy and capture the multielectron Coulomb interaction \cite{zajacScalableGateArchitecture2016}.



\subsection{Hubbard Parametrization}
We model the double quantum dot system using the extended Hubbard model without hopping derived from the capacitance model \cite{vanderwielElectronTransportDouble2002, yangGenericHubbardModel2011}. The Hamiltonian for this model is

\[
H \;=\; \sum_{i=1,2}\big(-\mu_i\,n_i \;+\; \frac{U_i}{2}n_{i}(n_{i}-1)\big) \;+\; U_{12}\,n_1 n_2
\]
where $\mu_i$ is the chemical potential of an electron at site $i$, $n_i$ is the electron number operator for site $i$, $U_i$ is the intra-site Coulomb interaction for site $i$, and $U_{12}$ is the inter-site Coulomb interaction between sites 1 and 2.  

We neglect tunnel coupling as it does not significantly alter charge stability diagrams. A constant tunnel coupling of $t\approx100\,\mu$eV and $t\approx300\,\mu$eV is added in appendix \ref{appendix_tunneling} to show it does not significantly alter results.
The main effect of this is the rounding of triple points \cite{yangGenericHubbardModel2011, wangQuantumTheoryChargestability2011,foulkTheoryChargeStability2024}. In addition, we neglect the spin-exchange, pair-hopping, and occupation-modulated hopping terms due to their minimal effect on the charge stability diagram. We define the lever arm matrix as \(L = \begin{bmatrix}\alpha_1 & \beta_1 \\ \beta_2 & \alpha_2\end{bmatrix}\), which converts a change in gate voltage to a change in dot energies. Certain models employ a normalization of the lever arms summing the diagonal and off diagonal components $\alpha + \beta = 1$ \cite{yangGenericHubbardModel2011, wangQuantumTheoryChargestability2011}. However, it is simple for us to directly calculate the lever arm as we already calculate the energies at each point in voltage space. Therefore, the calculated lever arms are a more appropriate choice.

We take constant values for $U_1, U_2,$ and $U_{12}$, as existing work has shown that these values do not vary significantly for modest gate voltage excursions \cite{merinoSimulatedChargeStability2024,foulkTheoryChargeStability2024, merinoSimulatedSpinQubits2025}. 
For more than one electron in the self-consistent SP loop, the intra-dot Coulomb interaction is included in the Poisson equation by subtracting the potential induced by the average charge density scaled to a charge of one electron. The resulting potential then approximately includes the intra-dot coulomb energy within the subdomains. 

\begin{figure}
      \centering
            \includegraphics[width=0.92\linewidth,trim=2.9cm 0.4cm 3.0cm 0.9cm,clip]{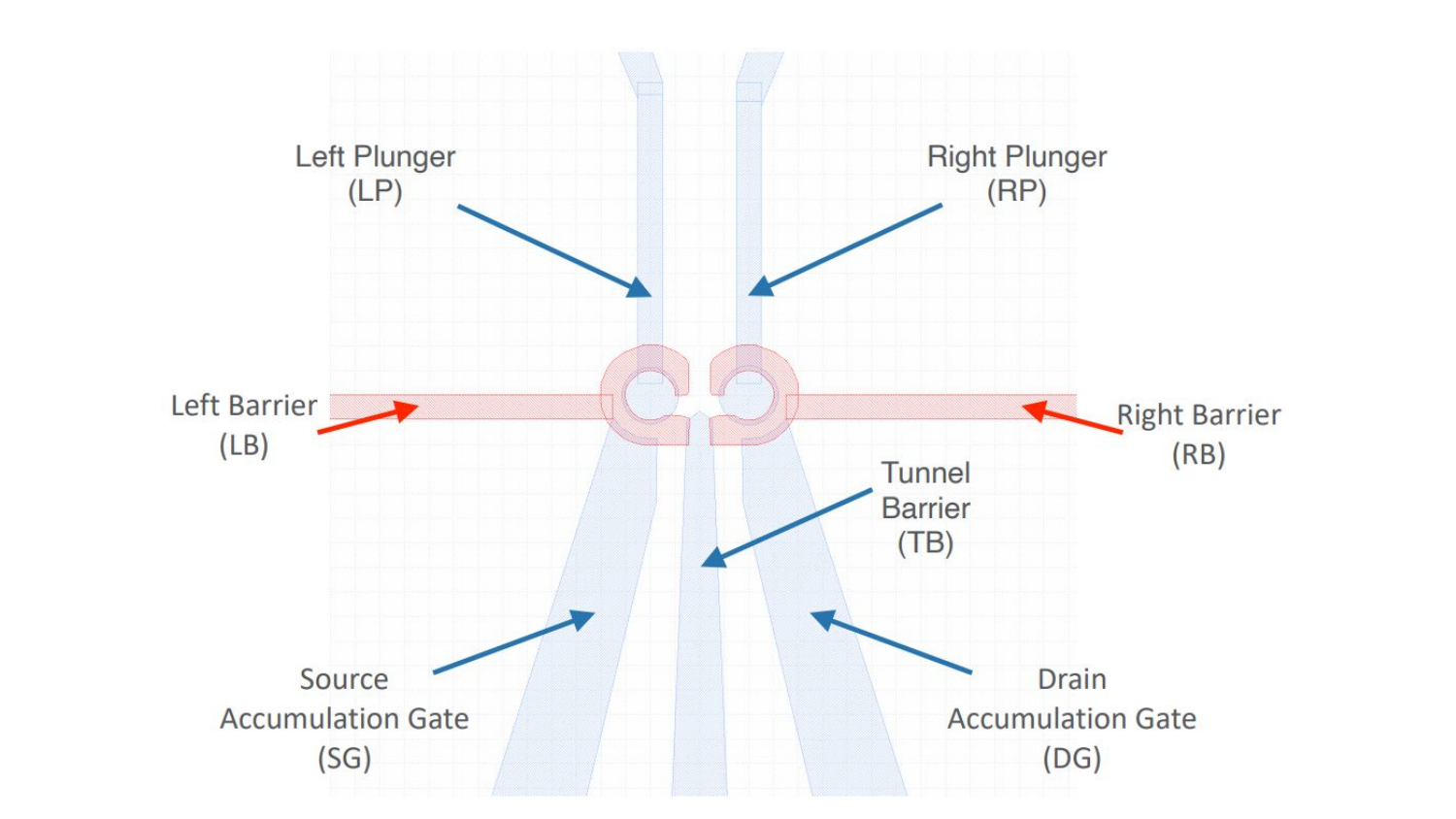}
      \caption{The schematic layout of the overlapping gate device design with each gate described. Additional information on this device is described in Section \ref{sec:results}.}
        \label{fig:tim_geometry}
\end{figure}

We focus on the charge configurations: (0,1), (1,1), (1,2), (2,1) and (2,2). For a symmetric device, we expect the eigenenergies for (1,2) and (2,1) to be equal, with $U_1=U_2$, $\alpha_1=\alpha_2$, and $\beta_1=\beta_2$. This is indeed the case for the symmetric second device in this paper, whose gate geometry is shown in Fig.~\ref{fig:tim_geometry}. The lever arms are calculated using a first order finite difference method to calculate the derivative of the energy of the electron states with respect to voltage. In the Fermi-Hubbard model, the chemical potentials are parameterized by the lever arms as
\[
\mu_1 = \alpha_1 V_1 + \beta_1 V_2 + \gamma_1 \label{mu_1}
\]
\[
\mu_2 = \beta_2 V_1 + \alpha_2 V_2 + \gamma_2\label{mu_2}
\]
where $\gamma_1$ and $\gamma_2$ are the constant energy shifts set by the potential in the device, and the lever arms are taken with the positive convention. The gate voltages $V_1$ and $V_2$ corresponding to each charge configurations are obtained via the MaSQE autotuning procedure. We apply this procedure at five different charge configurations, denoted by the number of electrons in P2 and P3: (0,1), (1,1), (1,2), (2,1), and (2,2). 
The lever arms are fixed across all configurations since they vary minimally in the low-charge regime as shown in appendix \ref{appendix_LAs}. We have also verified that the lever arm calculations do not vary significantly between single particle and FCI calculations in the (1,1) regime. For each configuration $(n_1,n_2)$ we equate the FCI ground-state energy computed at the autotuned voltages to the Hubbard energy of that occupation. Using these five configurations, we create a system of equations and solve for $U_1, U_2, U_{12}, \gamma_1$ and $\gamma_2$.

The fitted Hubbard model limited to two electrons per dot has a basis size of 16 states \cite{wangQuantumTheoryChargestability2011}. For each pair of chemical potentials in the left and right dot we diagonalize the Hamiltonian and identify its ground state using the Quantum Toolbox in Python (QuTiP). We then compute $\braket{n_1}$ and $\braket{n_2}$ for this ground state and round down to the nearest integer to assign the charge configuration at a pair of chemical potentials \cite{wangAutomatedCharacterizationDouble2023}. This process is analogous to and consistent with the thresholding routine conducted with MaSQE's continuous Thomas--Fermi charge filling model in the bias sweep simulations.

\subsection{Experimental Data}\label{Experimental_data}
\begin{figure*}[t]
    \centering
    \begin{minipage}[b]{0.284\textwidth}
      \centering
      \includegraphics[width=\linewidth]{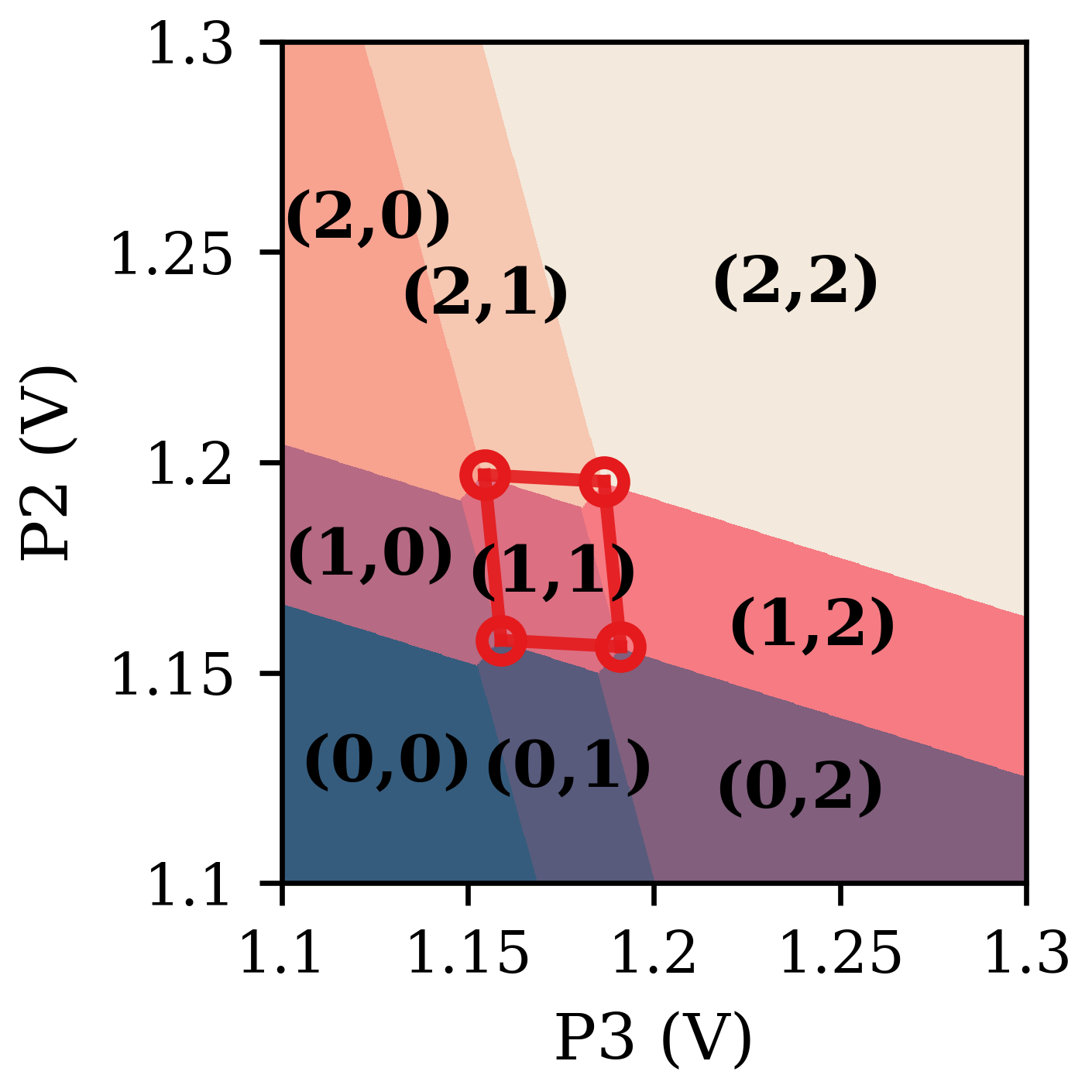}
      (a) FCI-Hubbard
    \end{minipage}
    \begin{minipage}[b]{0.284\textwidth}
      \centering
      \includegraphics[width=\linewidth]{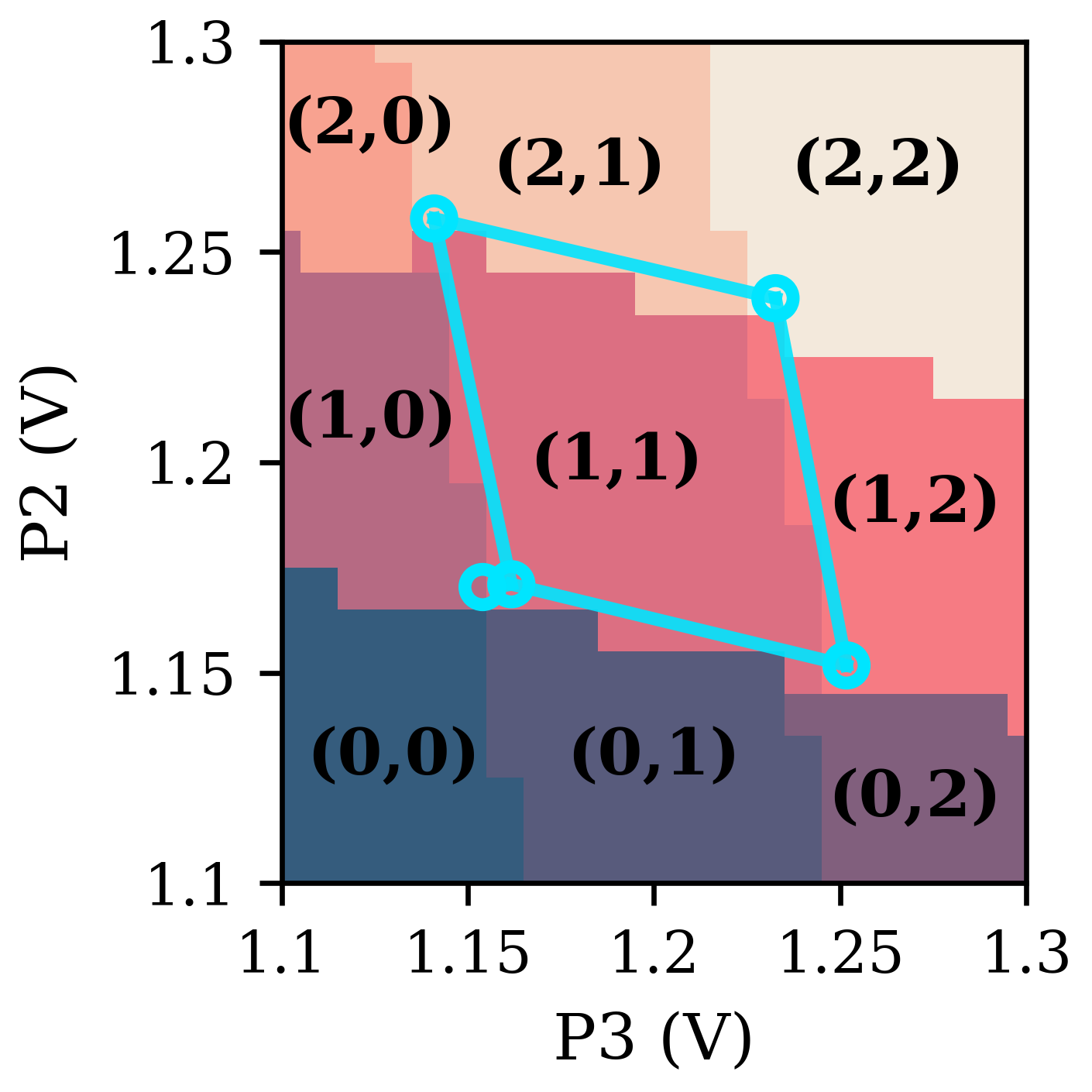}
      (b) Bias Sweep
    \end{minipage}
    \begin{minipage}[b]{0.394\textwidth}
      \centering
      \raisebox{-5.2pt}[\height][0pt]{\includegraphics[width=\linewidth]{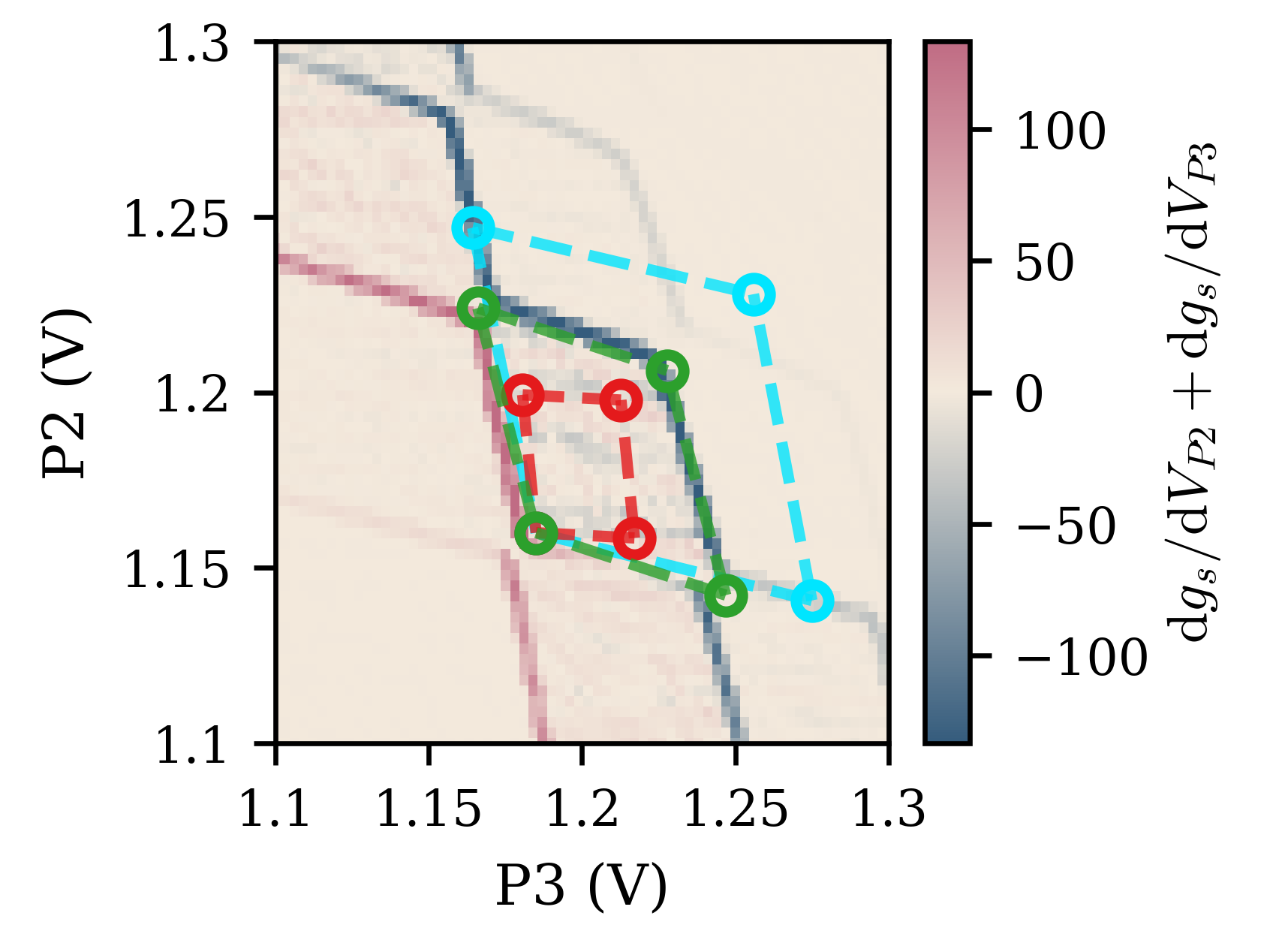}}
      (c) Experimental
  \end{minipage}
  \caption{The Intel device charge stability diagrams found using three methods. (a) A Hubbard model parameterized using five autotuned points indicated by blue circles in (b), with an FCI calculation at these points. This parameterized model was then used to generate the rest of the diagram, with the (1,1) charge cell displayed in red. (b) The charge stability diagram obtained by running self-consistent Schrödinger-Poisson simulations at 441 ($21\times21$) points in voltage space, with the autotuned points shown as reference. (c) The experimental data acquired using the procedure outlined in Sec. \ref{Experimental_data}. The 5~nm HfO$_2$ charging cells associated with the FCI-Hubbard for the (dashed red) and Autotuned Schrödinger-Poisson (dashed blue) are overlaid to illustrate deviation between the simulated and experimental data. With respect to the experimental data, the Hubbard model assuming 5~nm HfO$_2$ underestimates the charging energy, and the SP method overestimates the charging energy. The FCI-Hubbard model assuming 0~nm HfO$_2$ (dashed green) is closest to the measured cell. Voltage shifts to the simulated charge cells are chosen such that the $(1,1)$ onset point coincides with the experimental $(1,1)$ onset at $(V_{\mathrm{P3}}, V_{\mathrm{P2}}) \approx (1.185, 1.160)$ V.}
  \label{fig:Intel_trio}
\end{figure*}

Data are acquired from an Intel triple quantum dot (TQD) array using a dilution refrigerator with an electron temperature \(T_e = 100\) mK. The TQD device approximately follows the gate geometry shown in Fig.~\ref{fig:intel_1-1_geometry}. Charge sensing is performed by measuring the conductance, \(g_s\), of a charge sensor quantum dot located directly across from the TQD array. A double quantum dot (DQD) is formed in the triple quantum dot array by appropriately tuning the plunger and barrier gate voltages. The charge sensor is biased on the flank of a Coulomb blockade peak for sensitive charge detection. The DQD charge stability diagram data is acquired by measuring \(g_s\) as a function of the DQD plunger gate voltages. For visual clarity, we plot the summed derivative $\mathrm{d}g_s/\mathrm{d}V_{P2} + \mathrm{d}g_s/\mathrm{d}V_{P3}$, as shown in Fig.~\hyperref[fig:Intel_trio]{\ref{fig:Intel_trio}(c)}.


\section{Results}\label{sec:results}
In this section, we demonstrate the method for creating charge stability diagrams described in Section \ref{sec:Methods} on an Intel TQD device \cite{marcksValleySplittingCorrelations2025,yooDirectlyVisualizingEnergy2026,george12SpinQubitArraysFabricated2025, neyensProbingSingleElectrons2024,jandaMicrowaveResponseElectrically2026}, as well as an overlapping gate double dot claw shaped device \cite{wilsonFastSensitiveReadout2026}, shown in Fig.~\ref{fig:tim_geometry}. The generated charge stability diagrams from this method are then directly compared to experimental charge stability diagrams and extracted Hubbard parameters. 

\subsection{Intel Triple Quantum Dot}
The device gate geometry used is shown in Fig.~\ref{fig:intel_1-1_geometry}. As described in Section \ref{sec:Methods}, the plunger gate voltages of the five charge configurations, indicated by the blue circles in Fig.~\hyperref[fig:Intel_trio]{\ref{fig:Intel_trio}(b)}, are found using the autotune procedure and subsequently used to parameterize the model. For the purpose of our simulations, we initially assume the heterostructure from top to bottom consists of $\mathrm{HfO_2}$ (5~nm) $\rightarrow$ $\mathrm{SiO_2}$ (7.46~nm) $\rightarrow$ $\mathrm{Si}$ (1~nm) $\rightarrow$ $\mathrm{Si_{0.7}Ge_{0.3}}$ (50~nm) $\rightarrow$ $\mathrm{Si}$ quantum well (5~nm) $\rightarrow$ $\mathrm{Si_{0.7}Ge_{0.3}}$ (300~nm) \cite{george12SpinQubitArraysFabricated2025}. To understand how the Hubbard parameters change as a function of the dielectric used, we treat the HfO$_2$ thickness as an uncertain parameter and vary it below (5, 2, and 0 nm).

In order to probe the quantitative accuracy of our simulations, we use experimental voltages with a global voltage offset of $-1$ V. This offset corresponds to the accumulation gate voltage where the charge density is roughly $5 \times 10^{11}\ e/\text{cm}^{2}$ under the accumulation gates. Using a direct linear offset of all voltages, this offset sets the autotuned points closest to experimental values. The results are robust to a voltage offset between -0.95 and -1.05 volts, which gives rise to charge densities $4 \times 10^{11}\ e/cm^{2}$ to $6 \times 10^{11}\ e/cm^{2}$, demonstrating the capability of our method to explore different device regimes.



Table \ref{tab:gate_voltages} shows the gate voltages associated with an SP autotune for the five charge configurations of interest. At these points of interest, we conduct a 1, 2, 3, or 4 electron FCI calculation over the double dot system. Although five voltage sets are the minimum in solving for the five Hubbard parameters, this method could be improved when using a larger set of points, where each additional point sampled could result in a more precise Hubbard model.

We use a constant lever arm calculated at charge (1,1) which is consistent with FCI calculations as described in appendix \ref{appendix_LAs}.
The lever arm matrix we calculate for the (1,1) configuration is \(L = \begin{bmatrix} -0.096 & -0.027 \\ -0.029 & -0.093 \end{bmatrix}\) eV/V. These values are in line with the range of experimentally measured lever arms $\alpha_1$ and $\alpha_2$, which vary between $-0.070$ and $-0.100$ eV/V.

\begin{table}
\centering
\caption[Plunger Gate Voltages]{Plunger gate voltages $(V_1, V_2)$ at each charge configuration $(n_1, n_2)$, with $V_1 = V_{P2}$ (left dot) and $V_2 = V_{P3}$ (right dot).}
\label{tab:gate_voltages}
\begin{tabular}{|c|c|c|}
\hline
$(n_1, n_2)$ & $V_1$ (V) & $V_2$ (V) \\
\hline
$(0,1)$ & $1.170$ & $1.154$ \\
$(1,1)$ & $1.171$ & $1.162$ \\
$(1,2)$ & $1.152$ & $1.252$ \\
$(2,1)$ & $1.258$ & $1.141$ \\
$(2,2)$ & $1.239$ & $1.233$ \\
\hline
\end{tabular}
\end{table}

The resulting charge stability diagram calculated with the fitted Hubbard model is shown in Fig.~\hyperref[fig:Intel_trio]{\ref{fig:Intel_trio}(a)}. The Hubbard parameters found for these autotuned points are: $U_1 = 3.69$ meV, $U_2 = 2.98$ meV, $U_{12} = 0.72$ meV, $\gamma_1 = 142.3$ meV, and $\gamma_2 = 140.7$ meV. For comparison, Fig.~\hyperref[fig:Intel_trio]{\ref{fig:Intel_trio}(b)} shows a charge stability diagram where each point in voltage space corresponds to a self-consistent SP simulation. For this device, the five autotuned simulations and their FCI calculations used to parameterize the Hubbard model required a total of 32 simulations (27 self-consistent SP simulations with 5 FCI calculations) and approximately 1,251 minutes of computation time. In comparison, the fully self-consistent ($21\times21$) bias sweep required 441 self-consistent SP simulations and approximately 26,593 minutes. This corresponds to a reduction by a factor of $\sim14$ in the number of simulations and a factor of $\sim 21$ in total computation time. Of the 27 self-consistent SP simulations for the Hubbard model, two were used to construct the lever arm at the (1,1) configuration, which is reused for all chemical potentials. The disparity between methods would be even larger if we swept a more finely sampled or broader voltage space. This shows the advantage of using the autotuner to directly find the voltages of five specified charge occupations, instead of a self-consistent simulation for each voltage point to generate charge stability diagrams. Although the autotuned voltages do not lie exactly where the FCI-parametrized Hubbard model predicts the target occupation, they lie close to them. This matters because the FCI energies must be computed at a self-consistent potential near the correct occupation such that the Coulomb energies accurately represent the electrostatic environment the electrons experience in a device at that occupation.

The experimental charge stability diagram is shown in Fig.~\hyperref[fig:Intel_trio]{\ref{fig:Intel_trio}(c)}. For a 5 nm HfO$_2$, the dashed blue and dashed red lines indicate the shifted Schrödinger-Poisson and Hubbard model results, respectively. For a 0 nm HfO$_2$ (only 7.46 nm Si$O_2$), the Hubbard model results are overlaid in dashed green. Each is offset in voltage space to align with the (1,1) onset in the experimental data at $(V_{\mathrm{P3}}, V_{\mathrm{P2}}) \approx (1.185, 1.160)$~V (shifts of $23.3$ and $-11.0$ mV in P3 and P2 for the autotuned Schrödinger-Poisson points, $26.4$ and $2.6$ mV for the 5 nm HfO$_2$ Hubbard model onsets, and $-108.1$ and $-165.4$ mV for the 0 nm HfO$_2$ Hubbard model onsets), making the deviation in charging energy clearly visible. This is potentially a byproduct of uncaptured model disorder, such as trapped charges in the physical device, or due to the assumed dielectric stack.

Table \ref{tab:quantum_vs_exp} compares the Hubbard parameters fit from the five autotuned points against those extracted from experiment. The interdot charging energy $U_{12}$ and the offsets $\gamma_1$ and $\gamma_2$ fall within the experimental ranges, while the on-site charging energies are underestimated, with $U_1$ and $U_2$ falling $5.6\%$ and $16.5\%$ below the lower experimental bounds, respectively. We also attribute the underestimation of the charging energies to device disorder and mismatch between true device parameters and those used for simulation. The experimental Hubbard parameters are extracted directly from the experimental charge stability diagram by identifying four points on the charge-sensor derivative map: a $(0,0)$ reference and the $(1,1)$, $(1,2)$, and $(2,1)$ charge configurations. The on-site charging energies follow the plunger spacing of the cells as seen in \cite{wangAutomatedCharacterizationDouble2023}, converted from V to eV through the dot lever arms $\alpha_1$ and $\alpha_2$. The variation in measured lever arms directly translates to the uncertainty shown in Table \ref{tab:quantum_vs_exp}. Additional information on the experimental extraction can be seen in appendix \ref{appendix_extraction}.

\begin{table}[!htbp]
\centering
\caption[Simulation vs Experiment Hubbard Parameters]{Comparison of the simulated fit parameters assuming a 5 nm HfO$_2$ against experimental values. Ranges in experimental Hubbard parameters are a result of lever arm variation, with $\alpha_1$ and $\alpha_2$ observed between $-0.070$ and $-0.100$ eV/V.}
\label{tab:quantum_vs_exp}
\begin{tabular}{|c|c|c|c|c}
\hline
Parameter & Simulated & Experimental \\
\hline
$U_1$      & $3.69$ meV   & $3.91$--$5.58$ meV   \\
$U_2$      & $2.98$ meV   & $3.57$--$5.10$ meV   \\
$U_{12}$   & $0.72$ meV   & $0.61$--$0.88$ meV   \\
$\gamma_1$ & $142.3$ meV  & $104.1$--$148.6$ meV \\
$\gamma_2$ & $140.7$ meV  & $107.6$--$153.7$ meV \\
\hline
\end{tabular}
\end{table}

Another plausible contributor to the charging-energy discrepancy is the assumed dielectric stack. Our simulated routine was done using an assumed 5~nm HfO$_2$ thickness. Repeating the full autotune and FCI-Hubbard parametrization (with the lever arms recalculated at each thickness, see appendix~E) with a modified HfO$_2$ reduced to 2~nm shifts $U_1$ and $U_2$ significantly ($3.69 \rightarrow 4.20$~meV and $2.98 \rightarrow 3.86$~meV), moving both into the experimentally extracted range. However, the same re-tune lowers the fitted $U_{12}$ from 0.72 to 0.29~meV, outside the experimental range. Removing the HfO$_2$ entirely (0~nm), leaving only 7.46~nm SiO$_2$, raises $U_1$ and $U_2$ ($3.69 \rightarrow 5.18$~meV and $2.98 \rightarrow 5.15$~meV), placing $U_1$ within the experimentally extracted range and $U_2$ slightly above its upper bound, while the fitted $U_{12}$ falls from 0.72 to $-0.16$~meV, an unphysical attractive interdot interaction. 

Despite this, the 0~nm FCI-Hubbard cell is the closest in size to the measured $(1,1)$ cell, and can be seen in the green dashed cell in Fig.~\hyperref[fig:Intel_trio]{\ref{fig:Intel_trio}(c)}. We repeat this procedure in more detail, comparing the SP autotunes and FCI-Hubbard results at three different effective oxide thicknesses (5, 2, and 0~nm HfO$_2$) in appendix~\ref{appendix_oxide}. This investigation identifies that the Hubbard parameters mismatch cannot be purely attributed to oxide thickness, as no single HfO$_2$ thickness brings $U_1$, $U_2$, and $U_{12}$ into their experimental ranges simultaneously. Since we do not have access to direct experimental lever arms or material thicknesses, this limits our ability to precisely simulate the experimental Hubbard parameters. Inferences in this case are non-unique (comparable shifts could arise from variations in other heterostructure layers, quantum-well depth, or dielectric constants), but this exercise demonstrates how this framework can be inverted to constrain unknown material parameters from measured charge-stability data.

In addition to demonstrating the capabilities of Hubbard parametrization, this section demonstrates the computational strengths of an MDMM approach to SP. When utilizing an FCI calculation, the results we obtain via simulation show quantitative agreement with experimental data following an adjustment to the assumed dielectric stack. In the following section, we demonstrate how similar results in simulation are found for an overlapping gate device without using experimental voltages.

\subsection{Overlapping Gate Device}\label{novel_dev}
Having benchmarked the framework against an experimentally characterized charge stability diagram, we now apply it to a novel, overlapping gate double-dot device \cite{wilsonFastSensitiveReadout2026}. In contrast to the Intel Tunnel Falls
device, the charge stability diagram for this device was obscure in the low charge regime and is omitted. 
The autotuned operating points, lever arm matrices, and Hubbard parameters are obtained solely from the gate geometry and material heterostructure.

The Si/SiGe device, detailed and studied in \cite{wilsonFastSensitiveReadout2026} and shown in Fig.~\ref{fig:tim_geometry}, is focused on lever arm engineering that establishes scalable design principles for high-bandwidth dispersive readout. Prior to fabrication, the lever arms were optimized through gate design tests with MaSQE simulations. Simulated lever arms, \(L = \begin{bmatrix}-0.239 & -0.012 \\-0.012 & -0.239\end{bmatrix}\) eV/V, reasonably agreed with those measured for the fabricated device, \(L = \begin{bmatrix} -0.27 & -0.051 \\ -0.049 & -0.269 \end{bmatrix}\) eV/V \cite{wilsonFastSensitiveReadout2026}. 
The material stack from top to bottom was designed to be $\mathrm{Al_2O_3}$ (5~nm) $\rightarrow$ $\mathrm{Si}$ (2~nm) $\rightarrow$ $\mathrm{Si_{0.7}Ge_{0.3}}$ (50~nm) $\rightarrow$ $\mathrm{Si}$ quantum well (5~nm) $\rightarrow$ $\mathrm{Si_{0.7}Ge_{0.3}}$ (225~nm). Details on the material stack simulation methods can be found in appendix \ref{appendix_materials}.



The slightly larger values for the lever arm matrix elements seen in experiment as compared with the MaSQE simulation results are likely due to inconsistent atomic layer deposition of $\mathrm{Al_2O_3}$. 
The simulations are carried out with an oxide of 5 nm, but measurements of this for the experimentally realized device indicated it is roughly 3.5 nm. This signifies a key reason for the larger experimentally observed lever arm values as a thinner oxide increases the capacitive coupling of the gate to the well potential, thus increasing the lever arm. In this way, MaSQE captured behavior of the idealized device and helped identify a potential fabrication related source of discrepancy.

To quantify the sensitivity of the lever arms to the gate-oxide thickness, we repeat the lever arm calculations with the $\mathrm{Al_2O_3}$ thickness reduced from the nominal 5~nm to 3~nm. The resulting lever arm matrix is \(L_{3\,\mathrm{nm}} = \begin{bmatrix} -0.255 & -0.012 \\ -0.012 & -0.255 \end{bmatrix}\)~eV/V, corresponding to a $6.7\%$ increase in the diagonal lever arms for the thinner oxide and moving the simulated values toward those measured in experiment. The thinner oxide accounts for roughly half of the diagonal discrepancy, while the off-diagonal elements remain largely unchanged. The remaining deviation, particularly in the measured off-diagonal elements, suggests additional contributions not captured by oxide thickness alone, such as charge traps. This mirrors the oxide-thickness sensitivity identified for the Intel Tunnel Falls device, and demonstrates the use of the framework to guide lever arm engineering prior to fabrication.

\begin{figure}[t]
  \centering
  \begin{minipage}{0.23\textwidth}
    \centering
    \includegraphics[width=\linewidth]{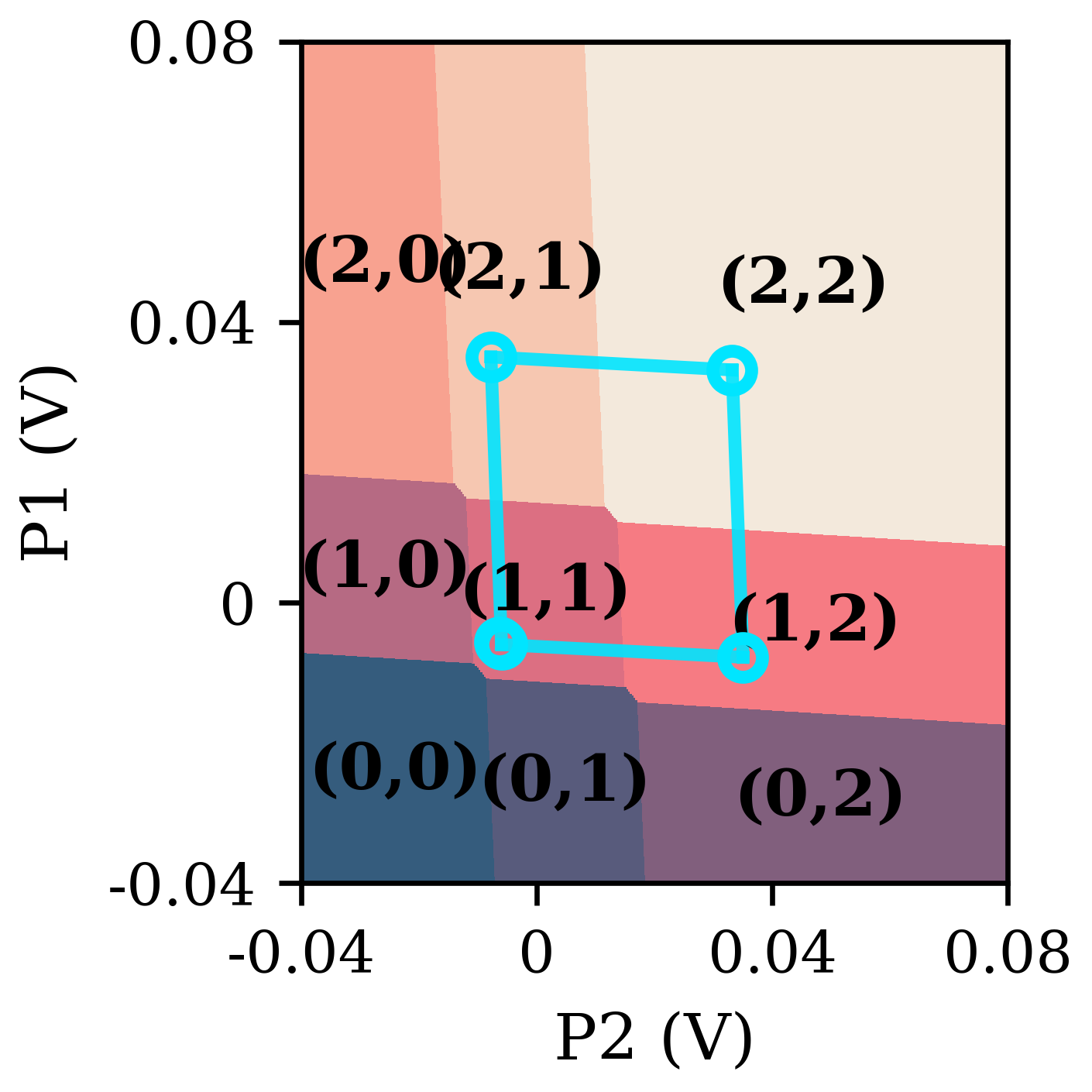}
    (a) Hubbard
  \end{minipage}
  \begin{minipage}{0.23\textwidth}
    \centering
    \includegraphics[width=\linewidth]{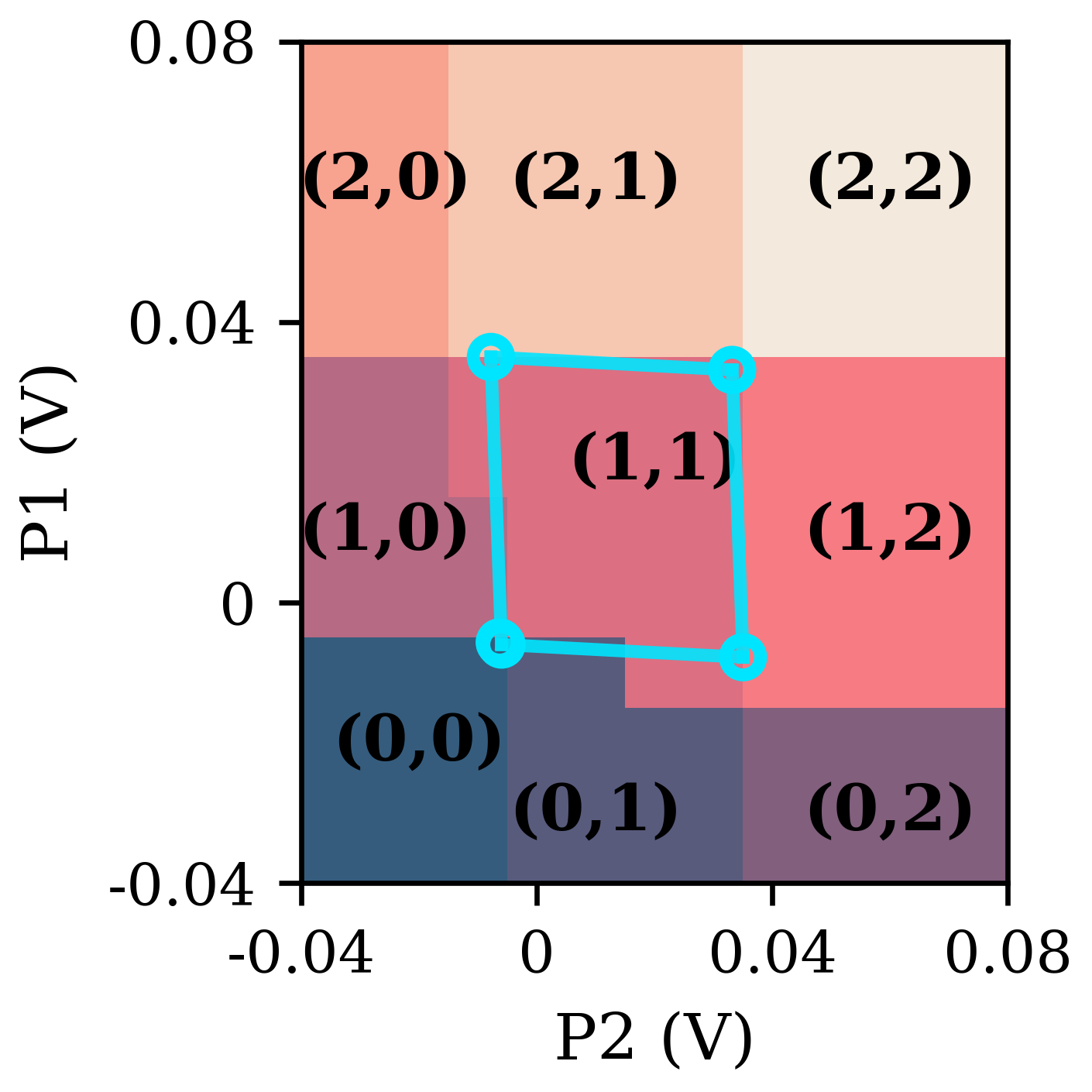}
(b) Bias Sweep
  \end{minipage}
  \caption{The overlapping gate device charge stability diagrams found using two simulation methods. (a) A Hubbard model parameterized using only five autotuned points (overlaid in blue) with an FCI calculation. This parameterized model was then used to generate the rest of the diagram. (b) The charge stability diagram obtained by running self-consistent Schrödinger-Poisson simulations at 169 ($13\times13$) points in voltage space, with the blue autotuned points shown as reference.}
  \label{fig:tim_duo}
\end{figure}

Figure \hyperref[fig:tim_duo]{\ref*{fig:tim_duo}(a)} shows the charge stability diagram generated from the fitted Hubbard model. The extracted parameters found are: \(U_1 = 6.11\) meV, \(U_2 = 6.10\) meV, \(U_{12} = -0.484\) meV, \(\gamma_1 = 2.20\) meV, and \(\gamma_2 = 2.19\) meV. 
The small negative fitted value of $U_{12}$ arises due to the relatively high barriers used (0.7 V plunger-barrier difference), and indicates minimal interdot interaction.
Figure \hyperref[fig:tim_duo]{\ref*{fig:tim_duo}(b)} shows a charge stability diagram where each point in voltage space corresponds to a self-consistent Schrödinger-Poisson simulation (using the semi-classical charge-filling model).  

For this device, the five autotuned simulations and their FCI calculations used to parameterize the Hubbard model required a total of 38 simulations (33 self-consistent SP simulations with 5 FCI calculations) and approximately 2,884 minutes of computation time. In comparison, the fully self-consistent ($13\times13$) bias sweep required 169 self-consistent SP simulations and approximately 40,439 minutes. This corresponds to a reduction by a factor of $\sim 4$ in the number of simulations and a factor of $\sim 14$ in total computation time. Of the 33 self-consistent SP simulations for the Hubbard model, two were used to construct the lever arm at the (1,1) configuration, which is reused for all chemical potentials. The disparity between methods would be significantly larger if we swept a larger or more finely sampled parameter space.

\section{Conclusion}\label{sec:conclusion}

We demonstrate a framework for rapidly generating charge stability diagrams in semiconductor quantum dot devices. By utilizing a multi-domain multi-model Schrödinger-Poisson methodology alongside an autotuning procedure, we identify gate voltages corresponding to selected charge configurations. To accurately capture the multielectron Coulomb interaction, we augment this autotuning procedure with a Full Configuration Interaction (FCI) calculation. From the resulting gate voltages, lever arms, and FCI energies, we extract the effective Hubbard parameters governing the double-dot system. This approach enables the construction of charge stability diagrams using fewer targeted simulations rather than performing a full voltage sweep. 

We validated this method on both an Intel Tunnel Falls Si/SiGe device and an overlapping gate device. For the Intel device, the initial simulated results agree qualitatively with the measured charge stability diagram. Quantitatively, the FCI-Hubbard model simulations underestimate charging energies, while the SP approach overestimates them. We then demonstrate how the intradot energy can enter anticipated experimental ranges through a modest correction to the assumed dielectric stack. Calculated lever arms reasonably agree with experimental expectations for both devices, and aligning the charge cells between simulation and experiment highlights the discrepancy in charging energy. These results demonstrate that the effects of reservoirs captured by the self-consistent Schrödinger-Poisson framework can be efficiently and effectively encoded in a Hubbard model.

This approach significantly reduces the computational cost of device level modeling while retaining the key physical features of the system. As a result, it provides a practical tool for the rapid exploration of device operating regimes, interpretation of experimental data, and pre-fabrication design of quantum dot architectures. In particular, the ability to quickly generate charge stability diagrams directly from the gate layout and heterostructure parameters moves toward the development of predictive digital twins for semiconductor quantum devices. 

Future investigations could include extending this framework by incorporating more accurate many-body energy calculations, such as a self-consistent FCI approach, to improve electron-electron interactions within dots or expand to characterizing exchange interactions. 
Additionally, future developments could also include using a more natural chemical potential model that aligns with Hubbard results, modeling device disorder, incorporating tunnel coupling when relevant, generating large datasets for machine learning approaches, and tuning larger devices with dozens of qubits.

\section{Acknowledgements}
This work used computational and storage services associated with the Hoffman2 Cluster which is operated by the UCLA Office of Advanced Research Computing’s Research Technology Group. The authors acknowledge support from the Army Research Office (ARO) under Grant Number W911NF-25-1-0141 and W911NF-23-1-0104. 

\bibliography{QubbardLink} 

\clearpage

\appendix
\onecolumngrid



\title{Appendix}
\maketitle

\section{Charge Model Choice}\label{appendix_full_3qd}
\begin{figure}[h]
        \centering
            \includegraphics[width=\linewidth]{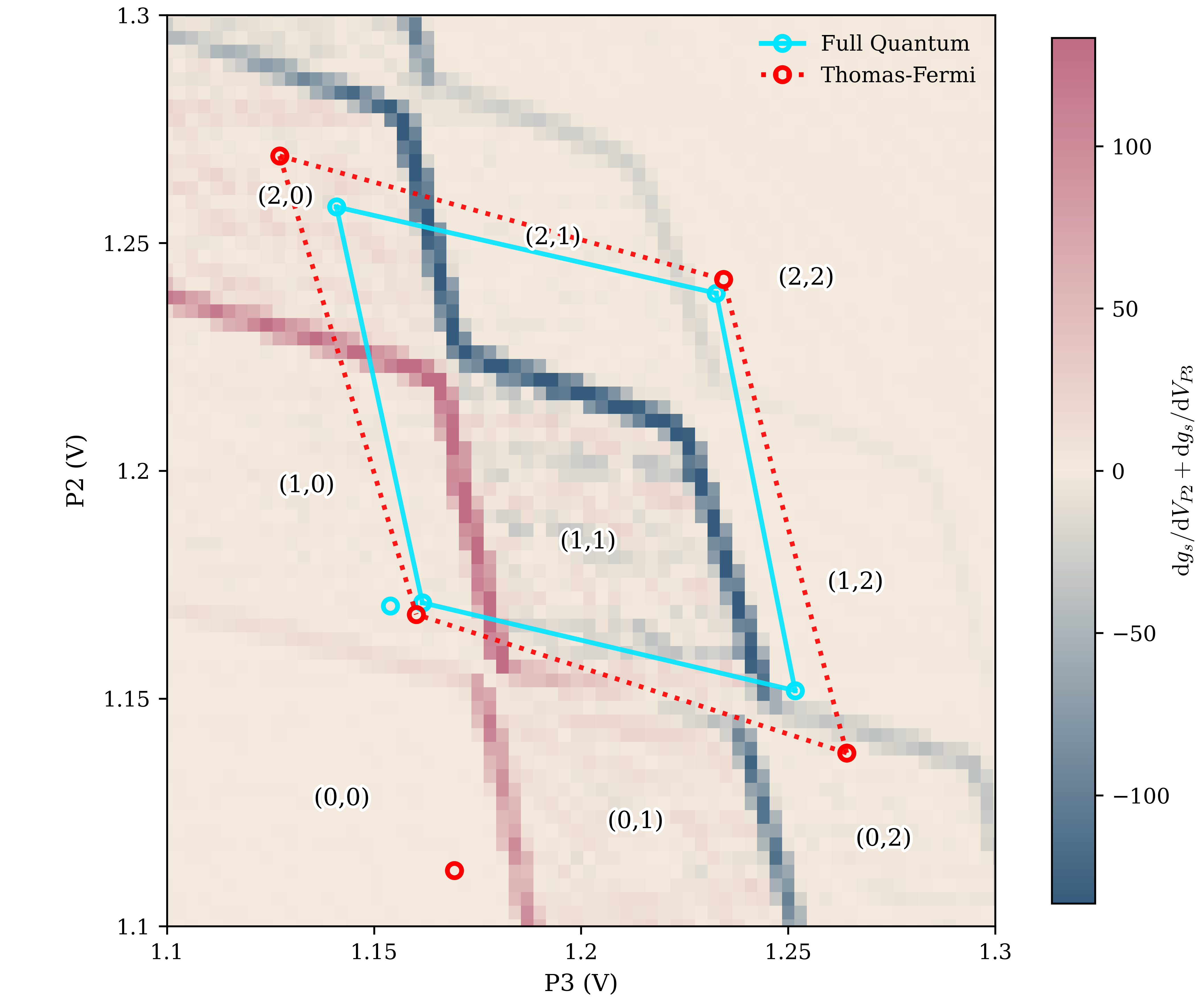}
        \caption[Charge Density Model Comparison]{Comparison of the Thomas--Fermi and fully quantum charge models. Two autotuning runs are overlaid: the semiclassical Thomas--Fermi model (red and dotted) and the quantum model (blue and solid). The Thomas--Fermi model overestimates the charging energies relative to the fully quantum model, and the fully quantum model overestimates charging energies relative to the experiment.}
        \label{fig:puddle_vs_quantum}
\end{figure}

We additionally benchmark the choice of charge model in the SP calculation. Fig. \ref{fig:puddle_vs_quantum} compares the autotuned operating points obtained from the semi-classical Thomas--Fermi model against those of the fully quantum model. Both runs are obtained using the same autotuning procedure, overlaid on the derivative of the experimental charge stability diagram. The Thomas--Fermi model deviates further from experiment and further overestimates the charging energies, motivating the use of the quantum model for all autotuned points throughout this work prior to doing FCI calculations. 

\section{Tunnel Coupling Effects}\label{appendix_tunneling}
\begin{figure}[h]
  \centering
  \begin{minipage}{0.30\textwidth}
    \centering
    \includegraphics[width=\linewidth]{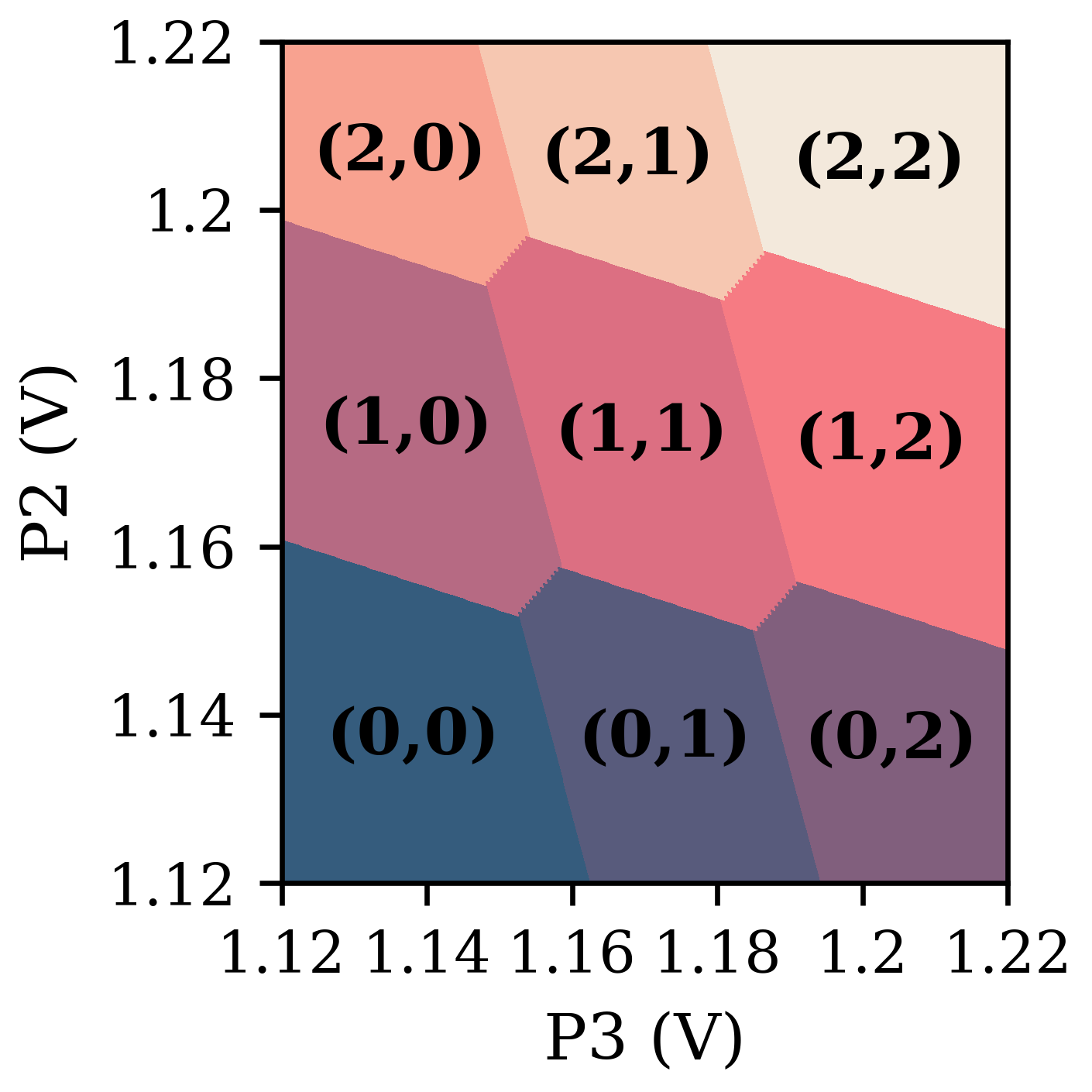}
    (a) $0\,\mu\mathrm{eV}$
  \end{minipage}
  \hfill
  \begin{minipage}{0.30\textwidth}
    \centering
    \includegraphics[width=\linewidth]{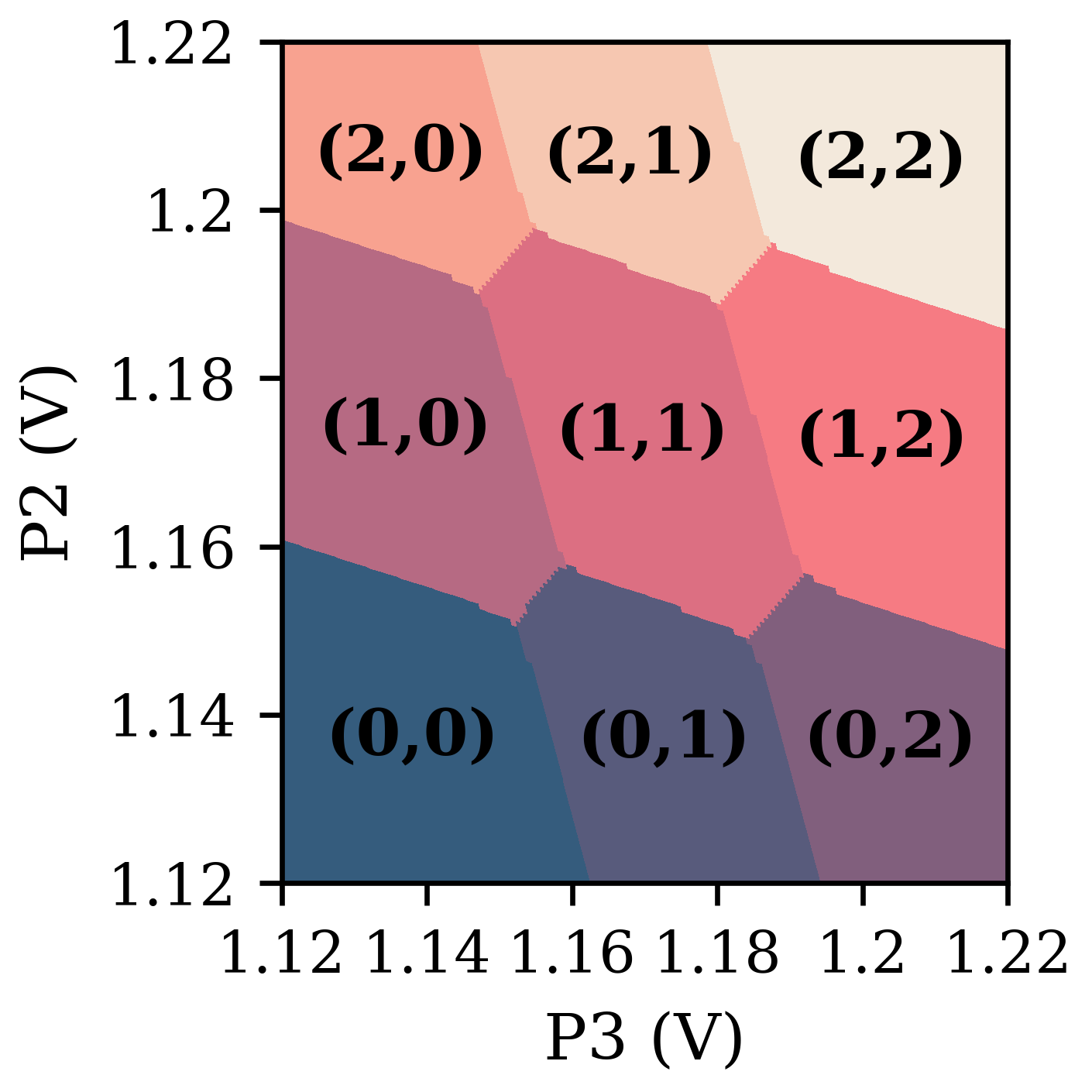}
    (b) $100\,\mu\mathrm{eV}$
  \end{minipage}
  \hfill
  \begin{minipage}{0.30\textwidth}
    \centering
    \includegraphics[width=\linewidth]{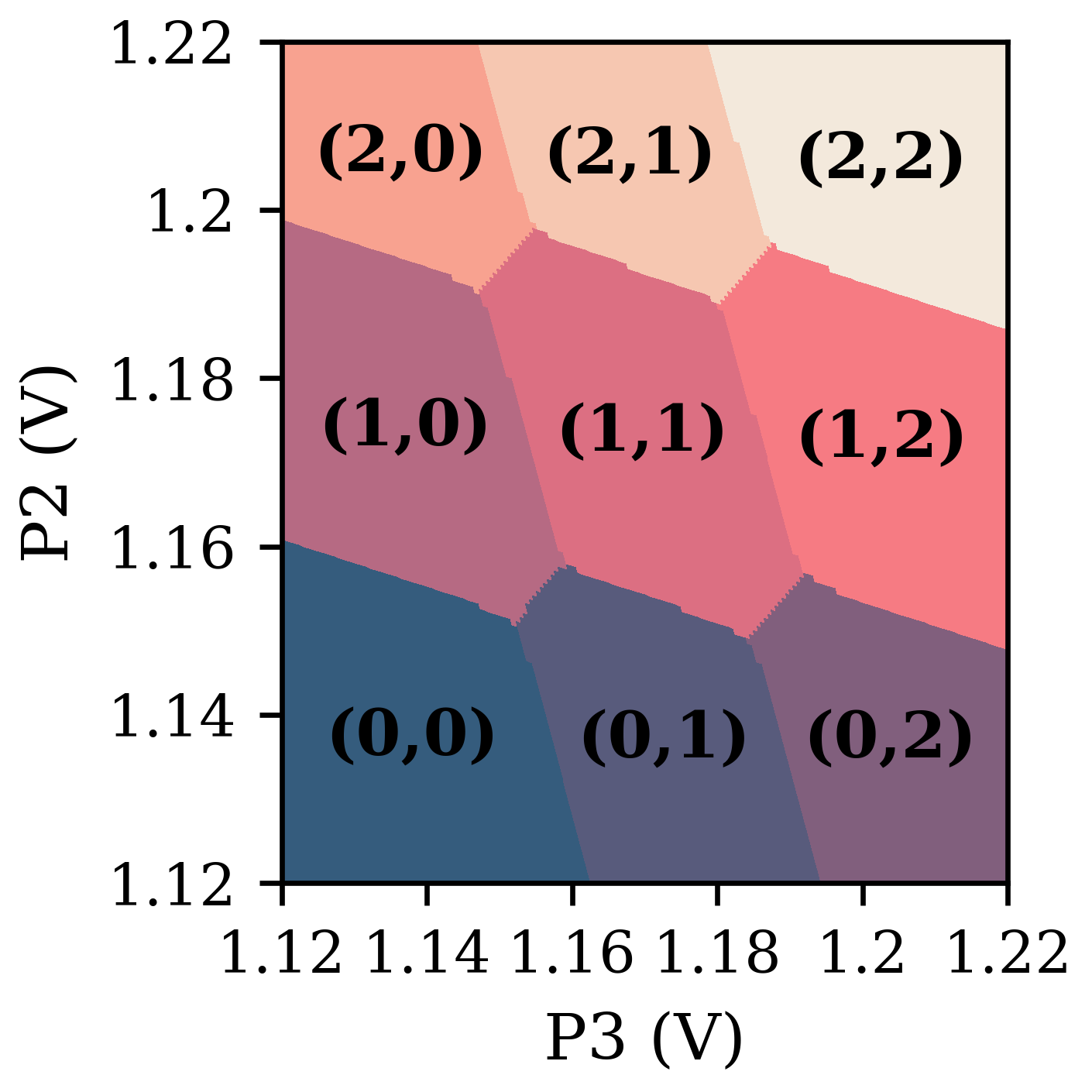}
    (c) $300\,\mu\mathrm{eV}$
  \end{minipage}
    \caption{Charge stability diagrams showing the effect of constant tunnel coupling with (a) $0\,\mu\mathrm{eV}$, (b) $100\,\mu\mathrm{eV}$, (c) $300\,\mu\mathrm{eV}$. The main effect is the rounding of triple points.}
    \label{fig:tunnel_coupling_rounding}
\end{figure}

In this section we examine the effects of tunnel coupling on the behavior of the charge stability diagram by comparing the Hubbard model diagram with and without this effect. Here, we extend the Hubbard model Hamiltonian to include the hopping term \[H_{t} \;=\; -t\sum_{\sigma=\uparrow,\downarrow}\Big(c_{1\sigma}^\dagger c_{2\sigma} \;+\; c_{2\sigma}^\dagger c_{1\sigma}\Big)\]
where $t$ is the tunnel coupling, $\sigma\in\{\uparrow,\downarrow\}$ indexes the spin, and $c_{i\sigma}$ is the annihilation operator of an electron at site $i$ with spin $\sigma$. 
The main effect of this is the rounding of triple points\cite{yangGenericHubbardModel2011, wangQuantumTheoryChargestability2011}, as we demonstrate in Fig. \ref{fig:tunnel_coupling_rounding} while other features do not change significantly. Realistically, the tunnel coupling changes across the voltage space of the charge stability diagram as a function of detuning \cite{merinoSimulatedSpinQubits2025}. For clarity, the following images use a constant tunnel coupling across chemical potential space in order to illustrate the rounding of triple points at every triple point.

\section{Lever Arm Variation}\label{appendix_LAs}
In this section, we optimize use of the linear model for lever arm matrix generation, as well as compare SP and FCI lever arm calculations. Generating a linear model at a given operating point requires two additional MDMM runs, in which the gate voltages are perturbed by $+0.001$ V and $-0.001$ V about the final autotuned point. Rather than regenerating a linear model at every autotuned charge configuration, we construct a single linear model at the $(1,1)$ configuration and reuse it for all configurations.

\begin{figure}
        \centering
            \includegraphics[width=\linewidth]{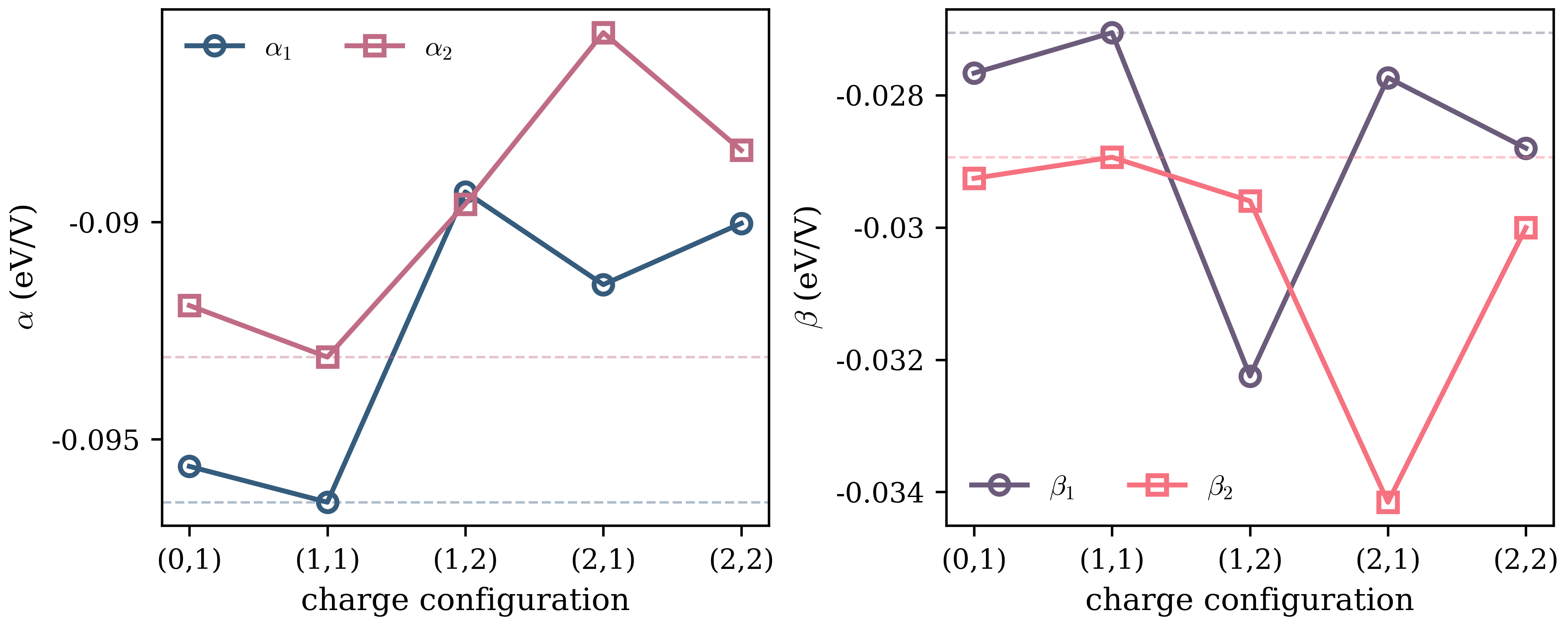}
        \caption[Lever Arm Variation Across Charge Configurations]{Variation in lever arms across autotuned points. A linear model is generated at each point via two additional MDMM runs ($\pm0.001$ V perturbations). The extracted lever arms vary minimally, justifying the reuse of a single linear model constructed at the $(1,1)$ configuration.}
        \label{fig:lever_arm_variation}
\end{figure}

The reference linear model, constructed at the $(1,1)$ configuration of the Intel device, yields the lever arm matrix \(L = \begin{bmatrix} -0.096 & -0.027 \\ -0.029 & -0.093 \end{bmatrix}\)~eV/V. Across the five charge configurations the diagonal lever arms deviate from their $(1,1)$ values by at most $8.0\%$ ($\alpha_1$: $7.4\%$, $\alpha_2$: $8.0\%$), and the off-diagonal lever arms by at most $19.2\%$ ($\beta_1$: $19.2\%$, $\beta_2$: $18.0\%$). Averaged over configurations, the lever arms are $\alpha_1 = -0.093 \pm 0.003$, $\beta_1 = -0.029 \pm 0.002$, $\beta_2 = -0.030 \pm 0.002$, and $\alpha_2 = -0.090 \pm 0.003$~eV/V (per-configuration values are tabulated in Table~\ref{app:lever_arms}). Since the off-diagonal elements are roughly a factor of three smaller than the diagonal ones, their larger relative variation has less effect on the reconstructed diagram, justifying the use of a single constant lever arm matrix evaluated at $(1,1)$ throughout this work. Fig. \ref{fig:lever_arm_variation} shows the lever arms extracted from linear models generated at each autotuned point, demonstrating that the lever arms vary only weakly across configurations and justifying the use of a single linear model. 

\begin{table}[h]
\centering
\caption[Lever Arms Across Charge Configurations]{Lever arm matrix elements calculated from linear models generated at each autotuned charge configuration. All values are in eV/V.}
\label{app:lever_arms}
\begin{tabular}{|c|c|c|c|c|}
\hline
Config & \(\alpha_1\) & \(\beta_1\) & \(\beta_2\) & \(\alpha_2\) \\
\hline
\((0,1)\) & $-0.096$ & $-0.028$ & $-0.029$ & $-0.092$ \\
\((1,1)\) & $-0.096$ & $-0.027$ & $-0.029$ & $-0.093$ \\
\((1,2)\) & $-0.089$ & $-0.032$ & $-0.030$ & $-0.090$ \\
\((2,1)\) & $-0.091$ & $-0.028$ & $-0.034$ & $-0.086$ \\
\((2,2)\) & $-0.090$ & $-0.029$ & $-0.030$ & $-0.088$ \\
\hline
\end{tabular}
\end{table}

As confirmation of the SP lever arm, we compute the (1,1) FCI ground-state energy at two gate voltages separated by $1\;\mathrm{mV}$ in P2. The ground-state energy moves from $-0.7360$ to $-0.6403\;\mathrm{meV}$, a shift of $0.0957\;\mathrm{meV}$ over $1\;\mathrm{mV}$, giving $|\alpha_1| = 0.0957\;\mathrm{eV/V}$. This agrees with the SP linear-model value of $0.09645$ eV/V at the $(1,1)$ configuration to within 1\%, confirming that the lever arms extracted from the SP model are consistent with the FCI calculation, and that both sit near the upper end of the experimentally observed window.

\section{Experimental Hubbard Parameter Extraction}\label{appendix_extraction}

\begin{figure}[h]
  \centering
  \includegraphics[width=\textwidth]{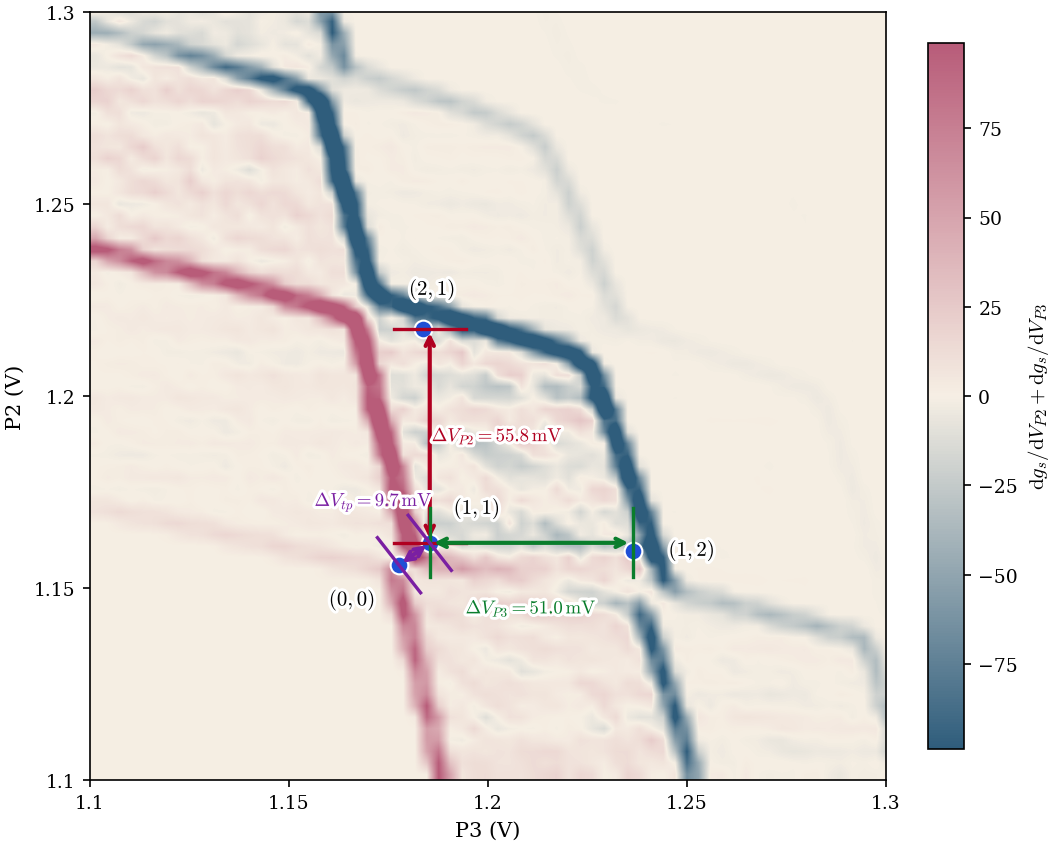}
  \caption[Experimental Hubbard Parameter Extraction]{Extraction of the Hubbard parameters from an experimental charge stability diagram. The charge-sensor derivative map ($\mathrm{d}g_s/\mathrm{d}V_{P2} + \mathrm{d}g_s/\mathrm{d}V_{P3}$) shows the honeycomb, with four identified points: a $(0,0)$ reference and the $(1,1)$, $(1,2)$, and $(2,1)$ configurations. The on-site energies $U_1 = \alpha_1\,\Delta V_{P2}$ and $U_2 = \alpha_2\,\Delta V_{P3}$ are read from the vertical and horizontal $(1,1)$ cell spacings, while the interdot coupling $U_{12}$ is obtained from the triple point splitting along the $(0,0)\!\to\!(1,1)$ diagonal, propagated through the lever arm matrix. Extracted values range over the lever arm window $|\alpha_{1,2}| \in [0.07, 0.10]$ eV/V: $U_1 = 3.91$--$5.58\;\mathrm{meV}$, $U_2 = 3.57$--$5.10\;\mathrm{meV}$, $U_{12} = 0.61$--$0.88\;\mathrm{meV}$, $\gamma_1 = 104.0$--$148.6\;\mathrm{meV}$, and $\gamma_2 = 107.6$--$153.7\;\mathrm{meV}$.}
  \label{fig:hubbard_experimental_extraction_2}
\end{figure}

Figure \ref{fig:hubbard_experimental_extraction_2} shows our extraction of Hubbard parameters from the experimental data of Fig.~\ref{fig:Intel_trio}(c), and shown in Table \ref{tab:quantum_vs_exp}. For the charging energies $U_1$ and $U_2$ we use the lever arm magnitudes, $\alpha_i = |\alpha_i|$. The dot-1 addition energy is read from the vertical $(1,1)\!\to\!(2,1)$ spacing, $U_1 = |\alpha_1|\Delta V_{P2}$, and the dot-2 addition energy from the horizontal $(1,1)\!\to\!(1,2)$ spacing \cite{wangAutomatedCharacterizationDouble2023}, $U_2 = |\alpha_2|\Delta V_{P3}$. Since the lever arms were observed in the experimental range $|\alpha_1|, |\alpha_2| \in [0.07, 0.10]$ eV/V, the spacings $\Delta V_{P2} = 55.8\;\mathrm{mV}$ and $\Delta V_{P3} = 51.0\;\mathrm{mV}$ give $U_1 = |\alpha_1|\Delta V_{P2} = 3.91$--$5.58\;\mathrm{meV}$ and $U_2 = |\alpha_2|\Delta V_{P3} = 3.57$--$5.10\;\mathrm{meV}$.

The inter-dot coupling $U_{12}$ is extracted from the triple-point splitting along the $(0,0)\!\to\!(1,1)$ diagonal. Unlike $U_1$ and $U_2$, this is not a single-gate displacement, so the voltage change in both gates must be propagated through the lever arm matrix,
\[
\Delta\mu_1 = \alpha_1\,\Delta V_{P2} + \beta_1\,\Delta V_{P3}
\]
\[
\Delta\mu_2 = \beta_2\,\Delta V_{P2} + \alpha_2\,\Delta V_{P3}
\]

The reference linear model, constructed at the $(1,1)$ configuration of the Intel device, yields the lever arm matrix  \(L_{\mathrm{sim}} = \begin{bmatrix} -0.096 & -0.027 \\ -0.029 & -0.093 \end{bmatrix}\) eV/V. Using the ratios $\beta_1/\alpha_1 = 0.28$ and $\beta_2/\alpha_2 = 0.31$ and corresponding the experimentally observed variation in $\alpha_i$, each $\beta_i$ can be approximated such that the full experimental matrix is \(L_{\mathrm{exp}} = \begin{bmatrix} -(0.07\text{--}0.10) & -(0.02\text{--}0.028) \\ -(0.022\text{--}0.031) & -(0.07\text{--}0.10) \end{bmatrix}\)eV/V.

We approximate that each chemical potential shift contributes equally to $U_{12}$:
\[
U_{12} = \tfrac{1}{2}\!\left(|\Delta\mu_1| + |\Delta\mu_2|\right)
\]

With the measured voltage change $(\Delta V_{P2}, \Delta V_{P3}) = (5.8,\,7.7)\;\mathrm{mV}$ (magnitude $\Delta V_{tp} = 9.7\;\mathrm{mV}$), the chemical-potential shifts range $|\Delta\mu_1| = 0.56$--$0.80\;\mathrm{meV}$ and $|\Delta\mu_2| = 0.67$--$0.95\;\mathrm{meV}$ across the lever arm window, giving $U_{12} = 0.61$--$0.87\;\mathrm{meV}$. As expected, $U_{12} \ll U_1, U_2$ for a weakly coupled double dot.

Finally, the energy offsets $\gamma_1$ and $\gamma_2$ are fixed by anchoring the model to where there are 0 electrons. We showed earlier how each chemical potential carries a constant offset on top of the gate contribution,
\[
\mu_1 = \alpha_1 V_{P2} + \beta_1 V_{P3} + \gamma_1
\]
\[
\mu_2 = \beta_2 V_{P2} + \alpha_2 V_{P3} + \gamma_2
\]
where the $\gamma_i$ absorb device voltage offset. At the $(0,0)$ top right corner, both dots sit on the verge of loading their first electron, so we set $\mu_1 = \mu_2 = 0$ there. Solving the two relations for the offsets then gives
\[
\begin{pmatrix}
\gamma_1 \\
\gamma_2
\end{pmatrix}
= -
\begin{pmatrix}
\alpha_1 & \beta_1 \\
\beta_2 & \alpha_2
\end{pmatrix}
\begin{pmatrix}
V_{P2} \\
V_{P3}
\end{pmatrix}
\]
with voltages at $(V_{P2}, V_{P3}) = (1.156, 1.178)$ V, we can solve for the gammas as $\gamma_1 = 104.0$--$148.6\;\mathrm{meV}$ and $\gamma_2 = 107.6$--$153.7\;\mathrm{meV}$ over the lever arm window. This offset shifts the assembled charge stability diagram to be at the correct reference point.

\section{Oxide Variation Results}\label{appendix_oxide}
The measured $(1,1)$ cell in Fig.~\hyperref[fig:Intel_trio]{\ref{fig:Intel_trio}(c)} spans $\Delta V_{P3} = 51.0$~mV and $\Delta V_{P2} = 55.8$~mV, which for lever arms between $-0.070$ and $-0.100$ eV/V corresponds to $U_1 = 3.91$--$5.58$~meV and $U_2 = 3.57$--$5.10$~meV as shown in Table~\ref{tab:quantum_vs_exp}. The Hubbard cell extracted for the initially assumed device stack (5~nm HfO$_2$) does not match the experimental (1,1) charge cell. The gate dielectric is unknown so we reduce it (2 and 0~nm HfO$_2$) and repeat the full simulation process: the autotuner locates the five charge configurations, the Hubbard parameters are solved using the FCI energies, and the charge onset corners are found.

\begin{figure}[t]
 \centering
 \includegraphics{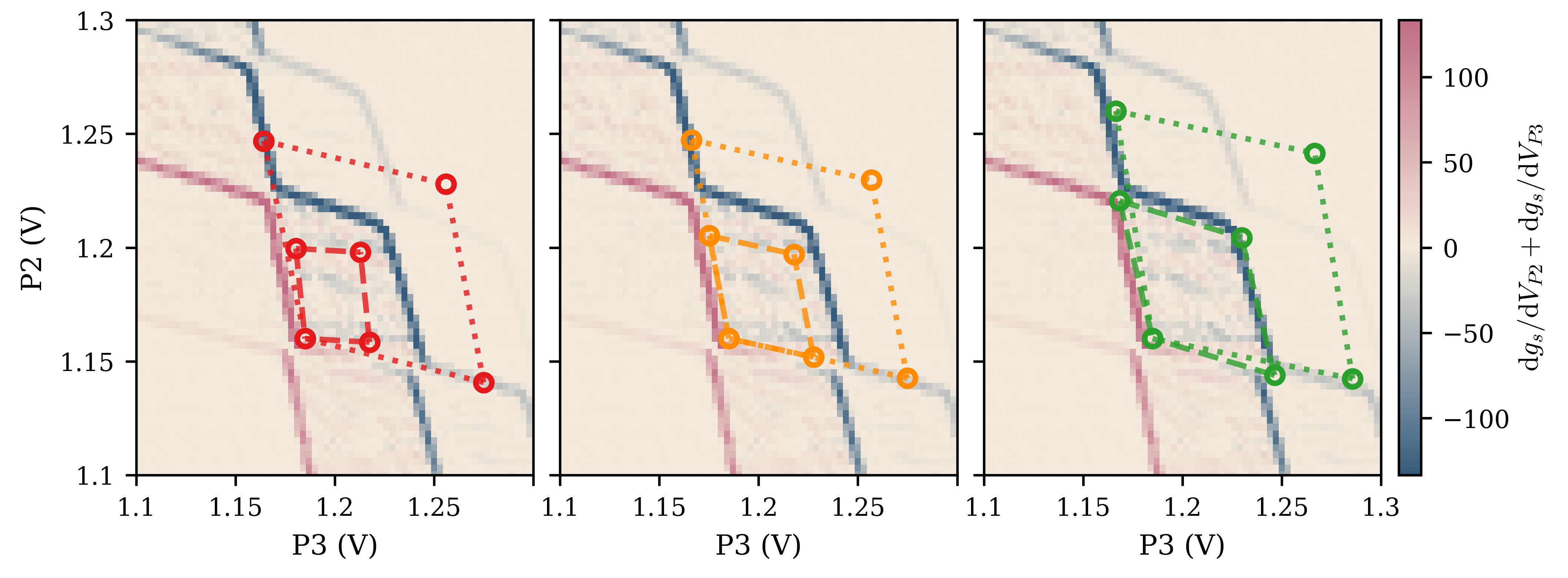}\\[-0.5ex]
 \makebox[7.057in][l]{\hspace*{1.507in}\makebox[0pt][c]{(a) 5 nm}\hspace*{1.907in}\makebox[0pt][c]{(b) 2 nm}\hspace*{1.907in}\makebox[0pt][c]{(c) 0 nm}}
 \caption{Simulated charge cell overlaid on the measured $\mathrm{d}g_s/\mathrm{d}V_{P2} + \mathrm{d}g_s/\mathrm{d}V_{P3}$ map for decreasing oxide dielectric (HfO$_2$ thickness of 5, 2, and 0~nm), (a) to (c), all over 7.46 nm of $\mathrm{SiO_2}$. Dotted lines connect the autotuned Schr\"odinger--Poisson configurations, dashed lines mark the FCI-Hubbard onset corners, and each cell is translated to the measured $(1,1)$ corner. The Hubbard cell grows toward the measured one as the dielectric is reduced, while the autotuned cell, already larger than experiment, grows further from it. This emphasizes how a change in oxide can affect the parameterized Hubbard model, and thus the charging energies.}
 \label{fig:oxide_series}
\end{figure}

Two cells are drawn per panel in Fig.~\hyperref[fig:oxide_series]{\ref{fig:oxide_series}}, with the autotuned Schr\"odinger--Poisson configurations (dotted), and the FCI-Hubbard onset corners (dashed). Both are translated so that their $(1,1)$ corner lands on the measured one at $(V_{\mathrm{P3}}, V_{\mathrm{P2}}) \approx (1.185, 1.160)$~V. The required shifts are small for the initially assumed 5 nm stack, $(+23.3, -11.0)$~mV for the autotuned points and $(+26.4, +2.6)$~mV for the Hubbard corners, and grow to $(-120.5, -177.1)$~mV and $(-108.1, -165.4)$~mV for the 0 nm stack in Fig.~\hyperref[fig:oxide_series]{\ref{fig:oxide_series}(c)}. This shift is a linear voltage offset, specified by $\gamma_{1,2}$ and the threshold of the effective stack, and does not enter the cell size or shape. The purpose of this comparison is in $U_1$, $U_2$, $U_{12}$, and the lever arms.

\begin{table}[!h]
\centering
\caption[Effective Dielectric Series Hubbard Parameters]{Hubbard parameters for the 5, 2, and 0~nm HfO$_2$ stacks of Fig.~\hyperref[fig:oxide_series]{\ref{fig:oxide_series}} against experimental values. Ranges in experimental Hubbard parameters are a result of lever arm variation, with $\alpha_1$ and $\alpha_2$ observed between $-0.070$ and $-0.100$ eV/V. Reducing the effective dielectric raises $U_1$ and $U_2$ but drives $U_{12}$ into an unphysical regime, as the inter-dot interaction should be positive/repulsive, yet (c)'s 0 nm HfO$_2$ FCI-Hubbard cell is the closest match to the experimentally measured cell.}
\label{tab:oxide_series}
\begin{tabular}{|c|c|c|c|c|}
\hline
Parameter & (a) 5~nm & (b) 2~nm & (c) 0~nm & Experimental \\\hline
$U_1$      & $3.69$ meV  & $4.20$ meV  & $5.18$ meV  & $3.91$--$5.58$ meV   \\
$U_2$      & $2.98$ meV  & $3.86$ meV  & $5.15$ meV  & $3.57$--$5.10$ meV   \\
$U_{12}$   & $0.72$ meV  & $0.29$ meV  & $-0.16$ meV & $0.61$--$0.88$ meV   \\
$\gamma_1$ & $142.3$ meV & $146.6$ meV & $149.2$ meV & $104.1$--$148.6$ meV \\
$\gamma_2$ & $140.7$ meV & $144.6$ meV & $145.1$ meV & $107.6$--$153.7$ meV \\
\hline
\end{tabular}
\end{table}

The lever arm matrices calculated at the (1,1) configuration for the three stacks are
\[
L_{5\,\mathrm{nm}} = \begin{bmatrix} -0.0965 & -0.0271 \\ -0.0289 & -0.0931 \end{bmatrix},\qquad
L_{2\,\mathrm{nm}} = \begin{bmatrix} -0.0985 & -0.0256 \\ -0.0271 & -0.0955 \end{bmatrix},\qquad
L_{0\,\mathrm{nm}} = \begin{bmatrix} -0.0915 & -0.0214 \\ -0.0219 & -0.0896 \end{bmatrix} \ \text{eV/V},
\]
all within the experimentally observed range. The diagonal elements change by ${\sim}5\%$ across the series while the off-diagonal elements fall by ${\sim}20\%$.

\section{Material Stack Simulation Parameters}\label{appendix_materials}

In this section we describe the material properties used in the simulation. For the Poisson equation the material properties used are the dielectric constants of the material layers, listed in Table \ref{tab:device_parameters}, along with an effective positive background doping of $10^{16}$ $e/\text{cm}^3$ caused by the growth process. The Schrödinger equation is solved using the effective masses and band offsets listed in Table~\ref{tab:device_parameters}. 

The boundary conditions for the Poisson equation are specified as follows. On the upper surface at points covered by metallic gates, values are set by the applied gate voltages, while at ungated surface points, values are determined through a domain decomposition procedure \cite{Anderson1989} 
where boundary values are determined so that a solution of Poisson's equation in the device with specified top surface boundary conditions couples consistently with a solution of Poisson's equation in the half-infinite region above the device. In the ungated regions on the top the potential is taken to approach zero far above the device. Periodic boundary conditions are applied in the lateral ($x$ and $y$) directions. At the bottom boundary, the potential is fixed to produce the charge density that compensates the background doping, as determined by our charge filling model. This requires the simulated heterostructure to extend sufficiently deep such that the Schrödinger-Poisson solution is converged with respect to bottom boundary depth.\par

\begin{table}[h]
\centering
\caption{Material parameters used in the device simulation.}
\label{tab:device_parameters}
\begin{tabular}{|l|c|c|c|c|}
\hline
Material & Dielectric Constant & \multicolumn{2}{c|}{Effective Mass ($m_0$)} & Band Shift \\
& $\epsilon_r$ & $m_{xy}$ & $m_z$ & $\Delta E$ (eV) \\
\hline
Al$_2$O$_3$        & $8.5$   &        &        &        \\
SiO$_2$            & $3.9$   &        &        &        \\
HfO$_2$            & $25$    &        &        &        \\
Si                 & $11.7$  & $0.199$  & $0.895$ & $-0.235$ \\
Si$_{0.7}$Ge$_{0.3}$ & $13.05$ & $0.1975$ & $0.916$ & $0$      \\
\hline
\end{tabular}
\end{table}

\end{document}